\documentclass[lettersize,journal]{IEEEtran}
\usepackage{amsmath,amsfonts,amsthm}
\usepackage{algorithmic}
\usepackage{algorithm}
\usepackage{array}
\usepackage[caption=false,font=normalsize,labelfont=sf,textfont=sf]{subfig}
\usepackage{textcomp}
\usepackage{stfloats}
\usepackage{url}
\usepackage{verbatim}
\usepackage{graphicx}
\usepackage{cite}
\usepackage{booktabs}
\usepackage{dcolumn}
\usepackage{tabularx}
 \usepackage{multirow}
\usepackage{color}
\usepackage{mathtools}

\usepackage[flushleft]{threeparttable}

\newenvironment{protocol}[1]
{\par\addvspace{\topsep}
\noindent
\tabularx{\linewidth}{@{} X @{}}
\hline
\textbf{#1} \\
\hline}
{ \\
\hline
\endtabularx
\par\addvspace{\topsep}}

\newcommand{\f}{\sf{f}}

\newtheorem{axiom*}{Axiom}
\newtheorem{assumption}{Assumption}
\newtheorem{theorem}{Theorem}
\newtheorem{lemma}[theorem]{Lemma}

\makeatletter
\newcommand{\leqnos}{\tagsleft@true\let\veqno\@@leqno}
\newcommand{\reqnos}{\tagsleft@false\let\veqno\@@eqno}
\reqnos
\makeatother

\makeatletter
\def\widebreve{\mathpalette\wide@breve}
\def\wide@breve#1#2{\sbox\z@{$#1#2$}%
     \mathop{\vbox{\m@th\ialign{##\crcr
\kern0.08em\brevefill#1{0.8\wd\z@}\crcr\noalign{\nointerlineskip}%
                    $\hss#1#2\hss$\crcr}}}\limits}
\def\brevefill#1#2{$\m@th\sbox\tw@{$#1($}%
  \hss\resizebox{#2}{\wd\tw@}{\rotatebox[origin=c]{90}{\upshape(}}\hss$}
\makeatletter

\newcommand{\UE}{\textit{UE}}
\newcommand{\SN}{\textit{SN}}
\newcommand{\HN}{\textit{HN}}

\begin{document}

\title{5G-AKA-HPQC: Hybrid Post-Quantum Cryptography Protocol for Quantum-Resilient 
\\5G Primary Authentication with Forward Secrecy}

\author{Primary authors: Yongho Ko, I Wayan Adi Juliawan Pawana, and Ilsun You}
%

\maketitle

\begin{abstract}
5G serves as a catalyst for transformative digital innovation by enabling convergence with various services in our daily lives. The success of this paradigm shift undeniably hinges on robust security measures, with primary authentication—securing access to the 5G network—being paramount. Two protocols, 5G Authentication and Key Agreement (5G-AKA) and the Extensible Authentication Protocol for Authentication and Key Agreement Prime (EAP-AKA'), have been standardized for this purpose, with the former designed for 3rd Generation Partnership Project (3GPP) devices and the latter for non-3GPP devices. However, recent studies have exposed vulnerabilities in the 5G-AKA protocol, rendering it susceptible to security breaches, including linkability attacks. Furthermore, the advent of quantum computing poses significant quantum threats, underscoring the urgent need for the adoption of quantum-resistant cryptographic mechanisms. Although post-quantum cryptography (PQC) is being standardized, the lack of real-world deployment limits its proven robustness. In contrast, conventional cryptographic schemes have demonstrated reliability over decades of practical application. To address this gap, the Internet Engineering Task Force (IETF) has initiated the standardization of hybrid PQC algorithms (HPQC), combining classical and quantum-resistant techniques. Consequently, ensuring forward secrecy and resilience to quantum threats in the 5G-AKA protocol is critical. To address these security challenges, we propose the 5G-AKA-HPQC protocol. Our protocol is designed to maintain compatibility with existing standard while enhancing security by combining keys negotiated via the Elliptic Curve Integrated Encryption Scheme (ECIES) with those derived from a PQC-Key Encapsulation Mechanism (KEM). To rigorously and comprehensively validate the security of 5G-AKA-HPQC, we employ formal verification tools such as SVO Logic and ProVerif. The results confirm the protocol's security and correctness. Furthermore, performance evaluations highlight the computational and communication overheads inherent to 5G-AKA-HPQC. This analysis demonstrates how the protocol effectively balances security and efficiency. In conclusion, our research provides significant insights into the design of secure, quantum-safe authentication protocols and lays the groundwork for the future standardization of secure authentication and key agreement protocols for mobile telecommunications.
\end{abstract}

\section{Preliminaries}
\subsection{Notation}
The explanation of symbols and terms used in the 5G-AKA-HPQC protocol is shown in Table \ref{5g-aka-hpqc_notation}
\begingroup

\begin{table}[h]
\begin{footnotesize}
\centering
\caption{Notations\label{5g-aka-hpqc_notation}}

\begin{tabularx}{\columnwidth}{|p{0.2\columnwidth}|p{0.7\columnwidth}|}
\hline
\textbf{Notation}	& \textbf{Meaning} \\
\hline
AK                              & Anonimity Key \\
AKA                             & Authentication and Key Agreement \\
AUSF                            & Authentication Server Function \\ 
AUTN                            & AUthentication TokeN \\ 
CK                              & Cipher Key \\ 
\(PK_{\HN}\)                    & HN's X-Wing encapsulation key \\
\(SK_{\HN}\)                    & HN's X-Wing decapsulation key \\
\((sk_{\UE}, pk_{\UE})\)        & UE's public-private key pair generated by X-Wing \\
\(PK_{\UE}\)                    & UE's public key computed by HN \\
eseed                           & expanded seed \\
\(ss_{\UE}\)                    & UE's shared secret generated by X-Wing encapsulation \\
\(c_{\UE}\)                     & UE's ciphertext generated by X-wing encapsulation \\
\(ss_{\HN}\)                    & HN's shared secret generated by X-Wing encapsulation \\
\(c_{\HN}\)                     & HN's ciphertext generated by X-wing encapsulation \\
\(k_{N}\)                       & shared key generated by N for SUCI encryption of decryption  \\
SUCI                            & SUbscriber Concealed Identifier \\
SUPI                            & SUbscription Permanent Identifier \\
\(SQN_{\HN}\)                   & HN's sequence number \\
\(SQN_{\SN}\)                   & SN's sequence number \\
RES                             & RESponse \\
XRES                            & eXpected RESponse \\
HPK                             & Hybrid Public Key \\
SEAF                            & SEcurity Anchor Function \\
SN                              & Serving Network \\
HN                              & Home Network \\
UE                              & User Equipment \\
KDF                             & Key Derivation Function \\
\(ID_{\SN}\)                    & Identifier of Serving Network \\
ECDH                            & Elliptic-curve Diffie–Hellman \\
MAC                             & Message Authentication Code \\
\hline

\end{tabularx}
\end{footnotesize}
\end{table}
\endgroup

\subsection{X-Wing}
X-Wing~\cite{b1} is a general-purpose post-quantum/traditional hybrid key encapsulation mechanism (PQ/T KEM) built on X25519~\cite{b2} and ML-KEM-768~\cite{b3} (NIST PQC level 1) for 128-bits security. The X-Wing has a triple of the following algorithms: Key generation, Encapsulation and Decapsulation.

\begin{footnotesize}
\begin{protocol}{X-Wing.KeyGen() - Key Generation}

    \textit{Inputs:} None \\
    \textit{The Protocol:}
    \begin{enumerate}
        \item Generate \(sk \leftarrow \text{random}(32)\)
        \item Calculate \(e \leftarrow \text{shake256}(sk, 96)\)
        \item \((sk1, pk1)\leftarrow\text{ML-KEM-768.KeyGen\_Internal}(e[0:32], \ e[32:64]) \)
        \item ECDH private key \(sk2 \leftarrow e[64:96]\)
        \item ECDH public key \(pk2 \leftarrow \text{X25519.DH}(sk2, gX25519) \)
        \item X-Wing public key \(pk \leftarrow (pk1, pk2) \)
        \item Return \((sk, pk)\)
    \end{enumerate}
    \textit{Outputs:} \(sk\)(secret key) and \(pk\)(public key)
    
\end{protocol}
\end{footnotesize}

    Key generation, returns 32 bytes secret key and 1216 bytes of public key. Secret key(sk) is used to decapsulate cipher text sent from the opponent and calculate shared secret. Public key(pk) is sent to the opponent and used to encapsulate the secret value. 

\begin{footnotesize}
\begin{protocol}{X-Wing.Enc($pk,~eseed$) - 
Encapsulation}
    \textit{Inputs:} X-Wing encapsulation key $pk$, (optional) X-Wing expanded seed $eseed$ \\
    \textit{The Protocol:}
    \begin{enumerate}
        \item{Parse $pk$ as $(pk1, pk2)$}
        \item{if $eseed$}
        \begin{enumerate}
            \item[2-1)]{Set $c2 \leftarrow eseed[32:64]$}
            \item[2-2)]{Compute $ss2 \leftarrow \text{X25519.DH}(eseed[0:32], ~pk2)$}
        \end{enumerate}
        \item{else}
        \begin{enumerate}
            \item[3-1)]{Generate \(ske \leftarrow \text{random}(32)\)}
            \item[3-2)]{Compute $c2 \leftarrow \text{X25519.DH}(ske, ~gX25519)$}
            \item[3-3)]{Compute $ss2 \leftarrow \text{X25519.DH}(ske, ~pk2)$}
        \end{enumerate}
        \item{$(ss1, c1) \leftarrow \text{ML-KEM-768.Enc}(pk1)$}
        \item{Set $s \leftarrow ss1 \parallel ss2 \parallel c2 \parallel pk2  \parallel \text{XWingLabel}\footnote{
            Hex value of XWingLabel is 5c2e2f2f5e5c. It represents the simbol X with 6 letter. (\texttt{\textbackslash./} and \texttt{/}\texttt{\^{}}\texttt{\textbackslash})
        } $}
        \item{Derive key $ss \leftarrow \text{SHA3-256}(s)$}
        \item{Set ciphertext $c \leftarrow (c1, ~c2)$}
        \item{Return ($ss, ~c$)}
    \end{enumerate}
    \textit{Outputs:} $ss$(shared secret), $c$(ciphertext)
\end{protocol}
\end{footnotesize}

The X-Wing encapsulation function used pk as an input and returns shared secret and cipher text. Inside the encapsulation process random ECDH key pairs are generated. This is probabilistic version of encapsulation function. However, for testing purpose, X-Wing specified deterministic version of Encapsulation function. Other then pk it uses 64 bytes $eseed$ as second input. $eseed$ will be used at the spot where it used to generate random value. With it, the encapsulation function becomes deterministic. To optimize the computation of proposed protocol 5G-AKA-HPQC, it was required to add deterministic option to the ECDH calculation. If the ECDH key generation and shared secret generation function were executed previously, with concatenating ECDH sk and ECDH pk as $eseed$ and use it as a second input of X-Wing encapsulation function. If $eseed$ exists on a input, encapsulation function will use $eseed$ as ECDH key and only calculate ECDH the shared secrete.

\begin{footnotesize}
\begin{protocol}{X-Wing.Dec($c, sk$) - Decapsulation}
    \textit{Inputs:} X-Wing ciphertext $c$, X-Wing decapsulation key $sk$ \\
    \textit{The Protocol:}
    \begin{enumerate}
        \item{Calculate $e$ $\leftarrow$ \text{shake256}($sk$, 96)}
        \item{($sk1,~pk1$) $\leftarrow$ \text{ML-KEM-768.KeyGen\_Internal}($e[0:32], e[32:64]$)}
        \item{ECDH private key $sk2$ $\leftarrow$ $e[64:96]$}
        \item{ECDH public key $pk2$ $\leftarrow$ \text{X25519.DH}($sk2, ~gX25519$)}
        \item{Parse $c$ as $(c1, ~c2)$}
        \item{ $ss1$ $\leftarrow$ \text{ML-KEM.Dec}($c1, sk1$)}
        \item{Compute $ss2$ $\leftarrow$ $\text{X25519.DH}$($sk_2, c2$)}
        \item{Set $s \leftarrow ss1 \parallel ss2 \parallel c2 \parallel pk2  \parallel \text{XWingLabel}$}
        \item{Derive key $ss \leftarrow \text{SHA3-256}(s)$}
        \item{Return $ss$}
    \end{enumerate}
    \textit{Outputs:} $ss$
\end{protocol}
\end{footnotesize}

Decapsulation function uses sk and the ciphertext sent from the opponent to calculate the X-Wing shared secret.

\section{Proposed Protocol}
\begin{figure}[h]

    \includegraphics[width=1.0\linewidth]{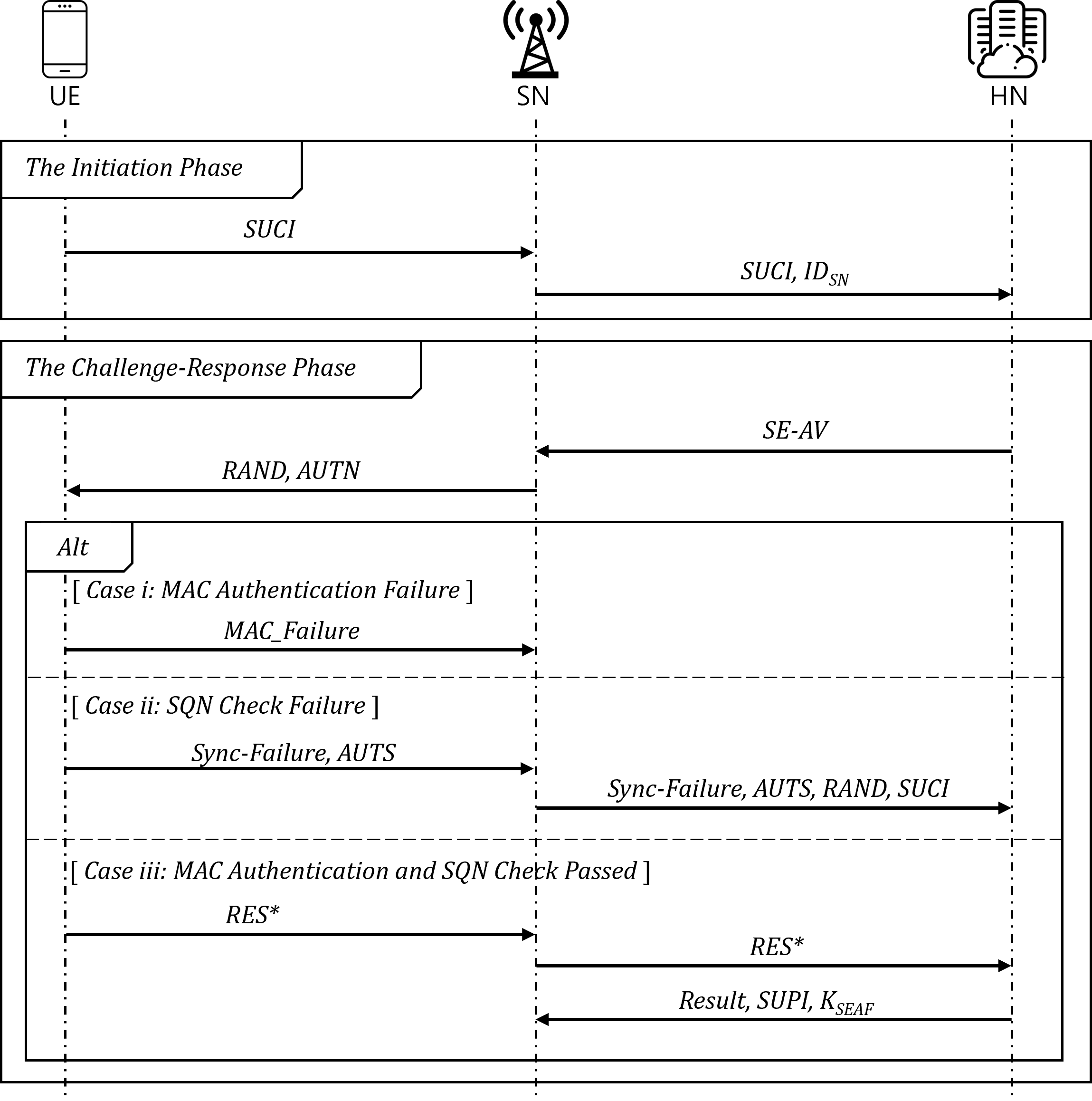}
    \caption{Proposed 5G-AKA-HPQC Protocol }
    \label{5G-AKA-HPQC Diagram}
\end{figure}

In this section, we present the 5G-AKA Hybrid Post-Quantum Cryptography (5G-AKA-HPQC) protocol, which combines classical cryptographic approaches with quantum-resistant methods to maintain compatibility with existing standardized protocols. The proposed protocol is illustrated in Figure \ref{5G-AKA-HPQC Diagram}.

\subsection{The Initiation Phase: Step 1}
\begin{footnotesize}
\begin{protocol}{Step 1.1 {(UE)}}
    \textit{Inputs:} With the HN's X-Wing encapsulation key $PK_{HN}$, UE executes the followings: \\
    \textit{The Protocol:}
    \begin{enumerate}
        \item \((sk_{UE},~pk_{UE}) \leftarrow \text{X-Wing.KeyGen}()\)
        \item Set \(eseed[0:32] \leftarrow \text{shake256}(sk_{UE}, 96)[64:96]\)
        \item Set \(eseed[32:64] \leftarrow \text pk_{UE}[1184:1216]\)
        \item \((ss_{UE}, ~c_{UE}) \leftarrow \text{X-Wing.Enc}(PK_{HN},~eseed)\)
        \item Compute $k_{UE} \leftarrow \text{KDF}\footnote{
            KDF function used in this step is ANSI-X9.63-KDF. With keyDataLength as 512 bits.
        }($ss$_{UE})$
        \item{Parse the $k_{UE}$ as ($k_1, k_2$)}
        \item{Parse $pk_{UE}~as~ (pk1_{UE},~ pk2_{UE})$}
        \item{Set $C_0 \leftarrow\ c_{UE}$}
        \item{Set $C_1 \leftarrow\ \text{ENC($k1$, $SUPI || pk1_{UE}$)}$}
        \item{Set $C_2 \leftarrow\ \text{MAC($k2$, $C_1$)}$}
        \item{Set $SUCI \leftarrow (C_0, C_1, C_2)$}
        \end{enumerate}
    \textit{Outputs:} UE sends $(SUCI)$ to SN.
\end{protocol}
\end{footnotesize}

    SUCI is a user privacy protection mechanism in 5G network. It uses public key of home network stored inside the USIM while generating fresh key pair and combine it to generate shared secret key to use as user identity(SUPI) encryption. By succeeding the architecture of traditional 5G, proposed protocol also assumes that the X-Wing pk of home network is stored inside the USIM. And by generating the fresh X-Wing key pair with X-Wing.KeyGen() function, it starts the 5G-AKA-HPQC protocol. In line 4 of Step 1.1, X-Wing.Enc() function is used to generate secret key to encrypt the SUCI and encapsulate the secret key to send it to the HN. However, like it were mentioned in X-Wing.Enc() explanation, the X-Wing.Enc() has ECDH key generation and ECDH shared secret generation process. Which has a redundancy with the X-Wing.KeyGen() function. To mitigate such redundancy and optimize the process, proposed protocol used extra input $eseed$, which is a combination of ECDH private key and ECDH public key generated in line 1. With generated secret key, UE will encrypt SUPI and it will generate tag using MAC for integrity verification.

\begin{footnotesize}
\begin{protocol}{Step 1.2 (SN)}
    \textit{Inputs:} SN receives $(SUCI)$ from UE. \\
    \textit{Outputs:} SN sends $(SUCI, ID_{SN})$ to HN.
\end{protocol}
\end{footnotesize}

\begin{footnotesize}
\begin{protocol}{Step 1.3 (HN)}
    \textit{Inputs:} Upon receiving  $(SUCI, ID_{SN})$ from SN, HN executes the following steps: \\
    \textit{The Protocol:}
    \begin{enumerate}
        \item Parse the $SUCI$ as $(C_0, C_1, C_2)$
        \item Decapsulate $\text{ss}_{\text{UE}} \leftarrow \text{X-Wing.Dec}(C_0, SK_{HN})$
        \item Compute $k_{\text{UE}} \leftarrow \text{KDF}(\text{ss}_{\text{UE}})$
        \item Parse the $k_{\text{UE}}$ as ($k_1 || k_2$)
    \end{enumerate}
    \textit{Outputs:} HN outputs $\bot$ if $C_2 \neq \text{MAC}(k_2, C_1)$. Otherwise, it outputs ($SUPI || \text{pk1}_{\text{UE}}) = \text{DEC}(k_1, C_1)$.
\end{protocol}
\end{footnotesize}

    SN will merely forward the SUCI to HN with SNN(Serving Network Name) attached in Step 1.2. HN will receive SUCI and SNN from the SN and by using X-Wing.Dec() function it will decapsulate the shared secret key to be used in SUCI integrity verification and decryption. 

\subsection{The Challenge-Response Phase: Step 2}

\begin{footnotesize}
\begin{protocol}{Step 2.1 (HN)}
\textit{Inputs:} Using $SQN_{HN}$, $k$ and shared secret ss sent from UE, HN generates an Authentication Vector $SE\text{-}AV = (RAND, AUTN, HXRES^*)$ as follows:\\
\textit{The Protocol:}
    \begin{enumerate}
        \item{Set $PK_{UE} \leftarrow\ (pk1_{UE}, C_0[1088:1120])$}
        \item{($ss_{HN}$, ~$c_{HN}$) $\leftarrow\ \text{X-Wing.Enc($PK_{UE}$)}$}
        \item Set RAND $\leftarrow c_{HN}$
        \item{Compute $HPK \leftarrow\ \text{KDF}\footnote{
            KDF function used in this step is ANSI-X9.63-KDF  with keyDataLength as 1120 bits. 
        }($ss$_{HN})$}
        \item{Compute $MAC \leftarrow {\f}_1 (k, SQN_{HN} || AMF || RAND \oplus HPK )$} 
        \item Compute $AK \leftarrow {\f}_5 (k, RAND \oplus HPK )$.
        \item Set $AUTN \leftarrow (AK \oplus SQN_{HN}, AMF, MAC)$
        \item Compute $CK \leftarrow {\f}_3 (k, RAND \oplus HPK )$
        \item Compute $IK \leftarrow {\f}_4 (k, RAND \oplus HPK )$
        \item Compute $XRES \leftarrow {\f}_2 (k,RAND \oplus HPK )$
        \item Compute $XRES^* \leftarrow {\sf KDF}\footnote{
            KDF function used in this step is HMAC-SHA256 with $CK || IK$ as a key.
        } (CK||IK, ID_{SN}||\newline RAND||XRES)$
        \item Compute $HXRES^* \leftarrow {LEFT}(128, {\text{SHA3-256}}\newline(RAND||XRES^*))$
        \item Derive $K_{AUSF} \leftarrow {\sf KDF}\footnote{
            KDF function used in this step is HMAC-SHA256 with $CK || IK$ as a key.
        }(CK||IK, ID_{SN}||AK \oplus SQN_{HN}||HPK)$ and \newline $K_{SEAF} \leftarrow {\sf KDF}\footnote{
            KDF function used in this step is HMAC-SHA256 with $K_{AUSF}$ as a key.
        } (K_{AUSF}, ID_{SN})$
        \item Increase $SQN_{HN}$ (i.e., $SQN_{HN} \leftarrow SQN_{HN} + 1$)
        \item Set $HE\text{-}AV \leftarrow (RAND, AUTN, XRES^*, K_{AUSF})$ 
        \item Set $SE\text{-}AV \leftarrow ( RAND, AUTN, HXRES^*)$
    \end{enumerate}
    \textit{Outputs:} HN sends $SE\text{-}AV$ to SN.
\end{protocol}
\end{footnotesize}

    In this phase, UE and HN authenticate each other via a challenge-response method and establish a quantum resistant and forward secrecy guaranteed anchor keys (i.e., $K_{SEAF}$) with SN. A core idea is that we set a \mbox{X-Wing} chphertext, which is sent through public channel, as a random challenge \textit{RAND}. And use X-Wing shared secret to generate $HPK$ and exclusive or to the RAND to use it to generate key materials. Since ecapsulation of shared secret used hybrid post quantum algorithm X-Wing, the attacker cannot access the shared secret even with the capability to break the traditional crypto algorithm ECDH. Also with shared secret protected with the public key of UE generated every authentication session, even with the long term key k and home networks private key, attacker cannot generate the anchor key. The detailed algorithm for the proposed protocol ars as follows.

    Among UEs X-Wing ciphertext $C_0$, last 32 bytes are ECDH public key. And decrypted value of SUCI includes pk1\_{UE}, which is ML-KEM public key of UE. With combination of these two values, PK\_{UE} is made. Using PK\_{UE} as a input of X-Wing.Enc() function, the HN generates shared secret for the challenge-response and anchor key. Since the input of the X-Wing.Enc() is only one, the pk of UE, it generates fresh ECDH keys inside the algorithm. The difference is, unlike the ECDH sk and pk of UE which should be used in a long term, the ECDH keys used in this phase only requires to be used once. With such reason, it can be temporally generated inside the algorithm without ever returning to store it.

    Since the freshly generated shared secret is shared in quantum resistant way and such shared secret is used as a input to generate key materials, this paper claims this protocol mutually authenticates UE and HN in a quantum resistant method. And also shares forward secrecy satisfying anchor key. This claim will be proved later in a section 3 with SVO logic~\cite{b4} and ProVerif~\cite{b5}.

\begin{enumerate}
\item[{\it Case i}]{(The SIM card returns $\bot$): The UE sends a failure message \texttt{Mac\_Failure} to the SN.}

\item[{\it Case ii}]{(The SIM card returns $AUTS$): The UE re-synchronizes with the HN by sending a failure message \texttt{Sync\_Failure} and $AUTS$ to the SN. Upon receiving the message $(\texttt{Sync\_Failure},$ $AUTS)$, the SN sends $(\texttt{Sync\_Failure}, ~AUTS, ~RAND, ~SUCI)$ to the HN. Then, the HN parses the received $AUTS$ as $(AK^* \oplus SQN_{UE}, AMF, MAC^*)$, and de-conceals $SQN_{UE}$} $\leftarrow AK^* \oplus SQN_{UE} \oplus {\f}_5^* (k, RAND \oplus HPK)$. Next, the HN checks its authenticity by comparing $MAC^* = {\f}_1^* (k, SQN_{UE} || AMF || RAND \oplus HPK)$. If the check holds, the HN re-sets $SQN_{HN}$ by $SQN_{UE} + 1$ (i.e., $SQN_{HN} \leftarrow SQN_{UE} + 1$).

\item[{\it Case iii}]{(The SIM card returns $(K_{SEAF}, RES^*)$): The UE stores $K_{SEAF}$ and sends $RES^*$ to the SN. Upon a recipient, the SN computes a hashed value \newline $HRES^* \leftarrow {\sf LEFT}(128, {\sf H}_{\text{SHA3-256}} (RAND||RES^*))$ and authenticates the UE by verifying if $HRES^* = HXRES^*$. If the comparison is positive, the SN forwards $RES^*$ to the HN. Next, the HN authenticates the UE by checking $RES^* = XRES^*$. If this comparison also holds, the HN sends its result along with $(SUPI, ~K_{SEAF})$ to the SN. The SN proceeds with the protocol only if both checks are successful; otherwise, it aborts the protocol. Once all checks pass, the two parties, UE and SN, communicate using session keys derived from the anchor key (i.e., $K_{SEAF}$) during subsequent 5G procedures. According to TS 33.501, both parties should implicitly confirm the agreed keys and each other's identities through the successful use of these keys in following procedures. This can be represented by a key-confirmation round trip using $K_{SEAF}$.}
\end{enumerate}

\begin{footnotesize}
\begin{protocol}{Step 2.2 (SN)} 
    \textit{Inputs:} SN receives $(SE\text{-}AV)$ from HN and stores $(RAND, HXRES^*)$. \\
    \textit{Outputs:} SN sends $(RAND, AUTN)$ to UE.
\end{protocol}
\end{footnotesize}

\begin{footnotesize}
    \begin{protocol}{Step 2.3 (UE)}
    \textit{Inputs:} Upon receiving $(RAND, AUTN)$ from SN, UE executes the following steps:
    \begin{enumerate}
        \item{\textcolor{black}{{In UE,} ME forwards the received $(RAND, AUTN)$ to SIM}}
        \item{Decapsulate $ss_{HN}$ ~ $\leftarrow\ \text{X-wing.Dec($RAND$, ~$sk_{UE}$)}$}
        \item{{Compute $HPK \leftarrow\ {\sf KDF}(ss_{HN})$}}
        \item{Run the SIM card command $\texttt{AUTHENTICATE} (RAND, HPK, AUTN)$}
        \begin{enumerate}
            \item[5-1)]{Compute $AK \leftarrow {\f}_5 (k, RAND \oplus HPK)$}
            \item[5-2)]{Parse $AUTN$ as $(CONC, AMF, MAC)$}
            \item[5-3)]{De-conceal $SQN_{HN} \leftarrow AK \oplus CONC$}
            \item[5-4)]{Check ${\f}_1 (k, SQN_{HN} || AMF || RAND \oplus HPK) = MAC$}
            \begin{itemize}
                \item If this check does not pass, the SIM card returns $\bot$ and then UE sends a failure message \texttt{Mac\_Failure} to SN (see \textbf{\textit{Case i}})
                \item{Otherwise, proceed to the next step}
            \end{itemize}
            \item[5-5)]{Check $SQN_{UE} < SQN_{HN} < SQN_{UE} + \Delta$~({The first condition $SQN_{UE} < SQN_{HN}$ ensures the} freshness of $(RAND,~AUTN)$. Also, the second condition $SQN_{HN} < SQN_{UE} + \Delta$, which is optional in the non-normative Annex C of TS 33.102, prevents a wrap-around of $SQN_{UE}$. {For example, if $\Delta$} is too small (i.e., $\Delta = 2$), an attacker can make a synchronization failure by sending $SUCI$ computed by the attacker with a fake SUPI. After this attack, the honest UE and HN can no longer authenticate each other. In TS 33.102, a recommended value of $\Delta$ is $2^{28}$ so as to decrease the synchronization failure rate.)}
            \begin{itemize}
                \item If this check does not pass, the SIM card computes \newline $MAC^* \leftarrow {\f}_1^* (k, SQN_{UE} || AMF || RAND \oplus HPK)$ and returns \newline $AUTS \leftarrow (AK^* \oplus SQN_{UE}, ~AMF, ~MAC^*)$ where $AK^* \leftarrow {\f}_5^* (k, RAND \oplus HPK)$. Then, UE re-synchronizes with HN by sending a failure message \texttt{Sync\_Failure} and $AUTS$ to SN (see \textbf{\textit{Case ii}})
                \item{Otherwise, proceed to the next step}
            \end{itemize}
            \item[5-6)]{Set $SQN_{UE} \leftarrow SQN_{HN}$}
            \item[5-7)]{Compute $CK \leftarrow {\f}_3 (k, RAND \oplus HPK)$ and $IK \leftarrow {\f}_4 (k, RAND \oplus HPK)$}
            \item[5-8)]{Compute $RES \leftarrow {\f}_2 (k, RAND \oplus HPK)~$, and \textcolor{black}{the SIM card returns $(RES, ~CK, ~IK)$ to the ME}}
        \end{enumerate}
        \item{\textcolor{black}{ME computes $RES^* \leftarrow {\sf KDF} (CK||IK, ~ID_{SN}||RAND||RES)$, and derives $K_{AUSF} \leftarrow {\sf KDF} (CK||IK, ~ID_{SN}||CONC||HPK)$ and $K_{SEAF} \leftarrow {\sf KDF} (K_{AUSF}, ID_{SN})$}}
        \item{\textcolor{black}{UE} returns $(K_{SEAF}, RES^*)$}
        \end{enumerate}
    \textit{Outputs:} UE stores $K_{SEAF}$ and sends $RES^*$ to SN (see \textbf{\textit{Case iii}})
\end{protocol}
\end{footnotesize}

\section{Formal Security Analysis}
\begin{figure}[H]
    
    \includegraphics[width=1.0\linewidth]{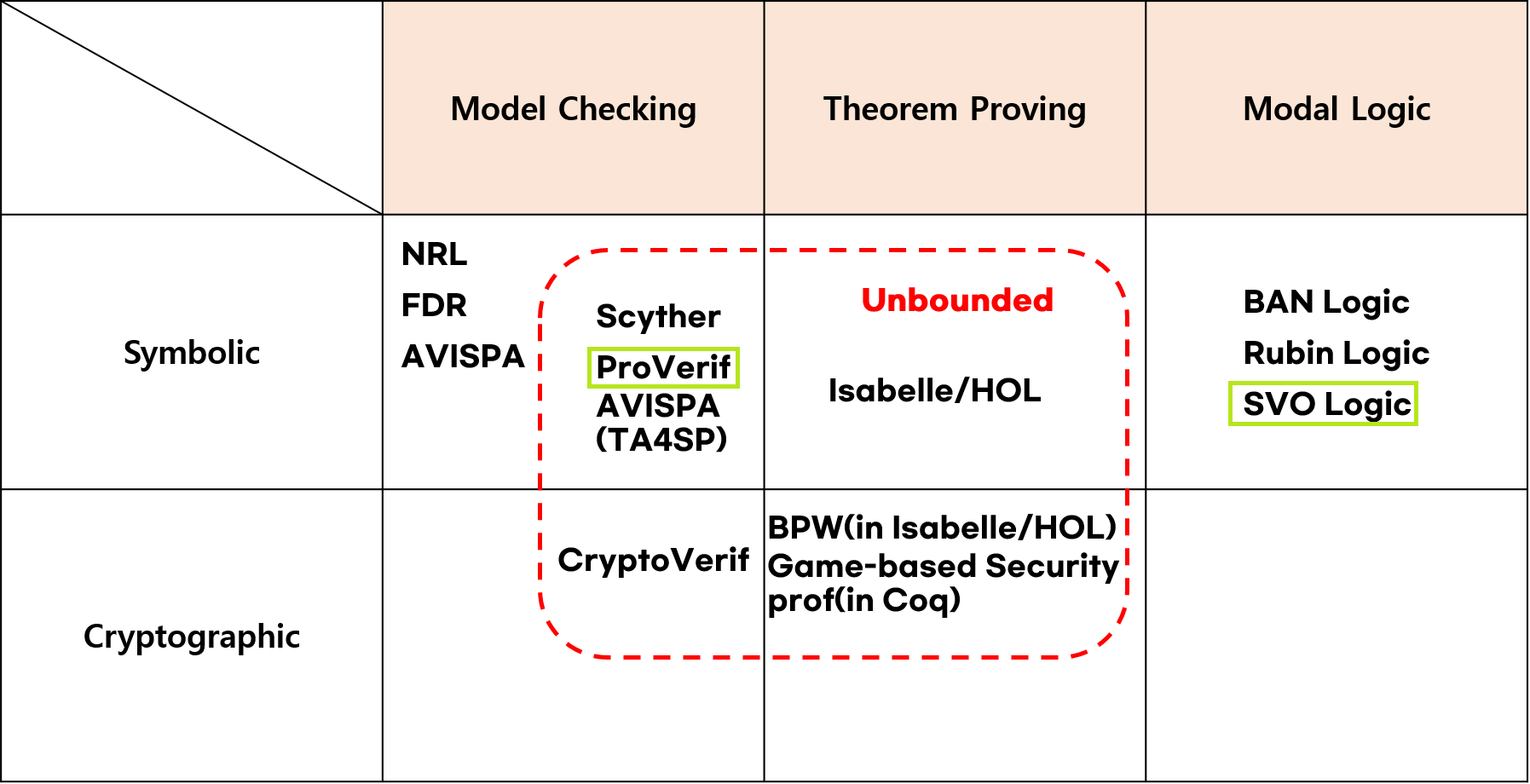}
    \caption{Formal verification categorization}
    \label{PV-all}
\end{figure}

In this section, we present a detailed and systematic security analysis of the 5G-AKA-HPQC Protocol using formal verification techniques. This analysis is conducted in alignment with the ISO/IEC 29128-1:2023 standard ~\cite{b6}, which establishes a rigorous framework for cryptographic protocol validation through mathematical proofs and vulnerability identification. The formal verification methods used are illustrated in Figure \ref{PV-all}, showcasing the integration of two advanced approaches: ProVerif and SVO Logic. These techniques represent distinct methodologies, each contributing uniquely to the comprehensive assessment of the protocol.

ProVerif, a state-of-the-art formal verification tool, employs an automated reasoning approach grounded in model checking principles. It is highly effective in analyzing cryptographic protocols by exploring all possible states and transitions within a system. This capability enables the detection of vulnerabilities and verification of security properties with exceptional scalability and performance. ProVerif’s unique strength lies in its ability to conduct "unbounded" verification, which goes beyond predefined constraints to assess a protocol's security in the most general scenarios. Additionally, ProVerif provides detailed attack trace graphs, facilitating the diagnosis and understanding of potential vulnerabilities and offering actionable insights for protocol improvement.

On the other hand, SVO Logic adopts a fundamentally different methodology, as it is rooted in modal logic. This advanced formalism is particularly suited for reasoning about security properties involving knowledge, trust, and belief. Unlike ProVerif’s model-checking approach, SVO Logic emphasizes rigorous logical analysis, enabling it to identify critical vulnerabilities with precision. Although SVO Logic does not support unbounded verification like ProVerif, it excels at detecting decisive flaws that might not be uncovered through model checking alone. Its focus on logical reasoning makes it a powerful complementary tool, providing an alternative dimension of analysis to strengthen the overall verification process.

By integrating these distinct methodologies, the security analysis achieves a multidimensional perspective, combining the exhaustive state exploration of ProVerif with the deep logical insights of SVO Logic. This dual approach ensures a robust and holistic evaluation of the 5G-AKA-HPQC Protocol, uncovering vulnerabilities from multiple angles and verifying its resilience against sophisticated adversaries. The complementary nature of these tools not only enhances the reliability of the assessment but also provides a blueprint for employing diverse formal verification techniques in future protocol analyses.

Through the integration of ProVerif’s model-checking capabilities and SVO Logic’s modal logic framework, this study underscores the importance of leveraging multiple methodologies for comprehensive cryptographic protocol evaluation. This combined approach ensures that the 5G-AKA-HPQC Protocol is rigorously tested, setting a high standard for security in next-generation network protocols.

\subsection{SVO Logic}

\begingroup
\renewcommand{\arraystretch}{1.2} 
\begin{table}[ht]
\centering
\caption{Notations of SVO Logic\label{svo logic notation}}
\begin{tabularx}{\columnwidth}{|p{0.3\columnwidth}|p{0.6\columnwidth}|}
\hline
\textbf{Notation}	& \textbf{Meaning} \\
\hline
\(\neg\) & Negation of formulae \\
\(P\ \textit{believes}\ X\)          & \(P\) believes the message \(X\) \\ 
\(P\ \textit{received}\ X\)          & \(P\) received the message \(X\) \\ 
\(P\ \textit{said}\ X\)              & \(P\) sent the message \(X\) \\ 
\(P\ \textit{says}\ X\)              & \(P\) sent the message \(X\) recently \\ 
\(P\ \textit{has}\ X\)               & \(P\) can see the message \(X\) \\
\(P\ \textit{control} \ X\)               & \(P\) has jurisdiction on \(X\) \\
\(\textit{fresh}(X)\)                 & The message \(X\) is fresh \\
\(PK_{\delta}(P,k)\)        & \(k\) is a public key-agreement key of \(P\) \\ 
\(\{X\}_{k}\)               & \(X\) is encrypted with \(k\) \\
\(\langle X \rangle_{*P} \) & \(X\) according to \(P\) \\
\(P \xleftrightarrow{k} Q\) & \(X\) is a secret key shared between \(P\) and \(Q\)\\
\hline
\end{tabularx}
\end{table}
\endgroup

\setlength{\belowdisplayskip}{4pt} \setlength{\belowdisplayshortskip}{4pt}
\setlength{\abovedisplayskip}{4pt} \setlength{\abovedisplayshortskip}{4pt}
\noindent 1. Belief Axiom (BA) 
\begin{equation} \notag
    (P\ \textit{believes}\ \varphi\ \wedge\ P\ \textit{believes} (\varphi \rightarrow \psi)) \rightarrow P\ \textit{believes}\ \psi
\end{equation}
\begin{equation} \notag
    P\ \textit{believes}\ \varphi \rightarrow \varphi 
\end{equation}
2. Source Association Axioms (SAA) 
\begin{multline*} \notag
    (P \xleftrightarrow{k} Q \wedge R\ \textit{received}\ \{X\ from\ Q\}_{k}) \rightarrow \\
    (Q\ \textit{said}\ X \wedge Q\ has\ X) 
\end{multline*}
3. Key Agreement Axiom (KA) 
\begin{equation}\notag
    (PK_{\sigma}(P,k_{P})\wedge PK_{\sigma}(Q,k_{Q})) \rightarrow P \xleftrightarrow{F_{0}(K_{P},K_{Q})} Q 
\end{equation}
4. Receiving Axioms (RA)
\begin{equation} \notag
    P\ \textit{received}\ (X_{1},...,X_{n}) \rightarrow P\ \textit{received}\ X_{i},\ for\ i= 1,...,n 
\end{equation}
\begin{equation} \notag
     P\ \textit{received}\ \{X_{k}\} \wedge P\ \textit{has}\ k^{-}) \rightarrow P\ \textit{received}\ X
\end{equation}
5. Possession Axioms (PA)
\begin{equation} \notag
    P\ \textit{received}\ X \rightarrow P\ \textit{has}\ X
\end{equation}
\begin{equation} \notag
    P\ \textit{has} (X_{1},...,X_{n}) \rightarrow P\ \textit{has}\ X_{i},\ for\ i= 1,...,n    
\end{equation}
6. Saying Axioms (SA)
\begin{equation} \notag
    P\ said\ (X_{1},...,X_{n}) \rightarrow P\ said\ X_{i} \wedge P\ has\ X_{i}
\end{equation}
\begin{multline*} 
    P\ \textit{says} (X_{1},...,X_{n}) \rightarrow \\ (P\ \textit{said} (X_{1},...,X_{n}) \wedge P\ \textit{says}\ X_{i}) \notag
\end{multline*}
7.  Freshness Axioms (FA)
\begin{equation} \notag
    \textit{fresh}(X_{i}) \rightarrow \textit{fresh}(X_{1},...,X_{n}),\ for\ i= 1,...,n
\end{equation}
\begin{equation} \notag
    \textit{fresh}(X_{1},...,X_{n}) \rightarrow \textit{fresh}\ F(X_{1},...,X_{n})
\end{equation}
8. Jurisdiction Axioms (JR) 
\begin{equation} \notag
    (P\ \textit{control}\ \varphi \wedge P\ \textit{says}\ \varphi) \rightarrow \varphi
\end{equation}
9. Nonce-Verification Axiom (NV)
\begin{equation} \notag
     (\textit{fresh}(X) \wedge P\ \textit{said}\ X) \rightarrow P\ \textit{says}\ X
\end{equation}

\subsection{Extension of SVO Logic}
We present the new axiom for the 5G-AKA-HPQC protocol, named the Perfect Forward Secrecy Axiom (PFSA). For this axiom to hold, two assumptions must be defined.

\begingroup

\renewcommand{\arraystretch}{1.3} 

\begin{table}[ht]
\caption{Perfect Forward Secrecy Notations\label{fps notation}}
\begin{tabularx}{\columnwidth}{|>{\centering}p{0.35\columnwidth}|p{0.55\columnwidth}|}
\hline
\textbf{Notation}	& \textbf{Meaning} \\
\hline
\(\textit{eph}(pk)\)                                     & \(pk\) is an ephemeral public key (or a set of ephemeral public keys) \\
\(\widebreve{PK}\)                              & \(\{pk_{1},...,pk_{n}\}\) \\
\( \textit{eph}(\widebreve{PK}) \)                       & \(eph(pk_{1}) \wedge ... \wedge eph(pk_{n})\) \\
\(own(P, eph(\widebreve{PK}))\)       & The subset of ephemeral public keys from the set \(eph(\widebreve{PK})\) owned by \(P\) \\
\(P\ allows\ \textit{fps}\Bigl(P \xleftrightarrow{K} Q\Bigl)\) & \(P\) trusts its own ephemeral public key(s), and although P has not gained sufficient trust in \(Q\)'s ephemeral public key(s), \(P\) allows that K satisfies Perfect Forward Secrecy \\
\hline
\end{tabularx}
\end{table}
\endgroup

\subsubsection*{\textbf{New Axiom}} 
We present the new axiom for the 5G-AKA-HPQC protocol, named the Perfect Forward Secrecy Axiom (PFSA). For this axiom to hold, two assumptions must be defined.
\begin{assumption} A set of ephemeral public keys \(eph(\widebreve{PK})\) is union of the subset of ephemeral public keys from the set\ \(eph(\widebreve{PK})\) owned by \(P\) and the subset of ephemeral public keys from the set\ \(eph(\widebreve{PK})\) owned by \(Q\), denoted as 
\begin{equation}
    eph(\widebreve{PK})\ =\ own\Bigl(P,eph(\widebreve{PK})\Bigl)\ \cup\ own\Bigl(Q,eph(\widebreve{PK})\Bigl) \notag
\end{equation}
\end{assumption}
\begin{assumption} The subset of ephemeral public keys from the set\ \(eph(\widebreve{PK})\) owned by \(P\) and the subset of ephemeral public keys from the set\ \(eph(\widebreve{PK})\) owned by \(Q\) are mutually exclusive, denoted as
\begin{equation}
    own\Bigl(P,eph(\widebreve{PK})\Bigl)\ \cap\ own\Bigl(Q,eph(\widebreve{PK})\Bigl) = \emptyset \notag
\end{equation}
\end{assumption}

Once the two assumptions are satisfied, we can proceed to define the perfect forward secrecy axiom.

\begin{axiom*}[Perfect Forward Secrecy] The Perfect Forward Secrecy Axiom (PFSA) is designed to account for two cases. For a set of public keys \( \widebreve{PK} \), if \(|\widebreve{PK}| > 1 \) then, 
\begin{multline*} \tag{i}\label{eq:fps:i}
    \textit{own}\Bigl(P, \textit{eph}(\widebreve{PK})\Bigl) \wedge\ Q\ \textit{says}\ \textit{own}\Bigl(Q,\textit{eph}(\widebreve{PK})\Bigl)\ \rightarrow \\
    P\ \textit{allows}\ \textit{fps}\Bigl(P\xleftrightarrow{F_{0}(\widebreve{PK})}Q\Bigl) 
\end{multline*}
\begin{multline*} \tag{ii}\label{eq:fps:ii}
    \textit{own}\Bigl(Q, \textit{eph}(\widebreve{PK})\Bigl) \wedge\ P\ \textit{says}\ \textit{own}\Bigl(P,\textit{eph}(\widebreve{PK})\Bigl)\ \rightarrow \\
    Q\ \textit{allows}\ \textit{fps}\Bigl(P\xleftrightarrow{F_{0}(\widebreve{PK})}Q\Bigl) 
\end{multline*}
if the condition \eqref{eq:fps:i} and \eqref{eq:fps:ii} hold, then the key \(P\xleftrightarrow{F_{0}(\widebreve{PK})}Q\) satisfies Perfect Forward Secrecy. \\

On the other hand, if the set \(\widebreve{PK}\) contains only one element \(pk\), then 
\begin{multline*} \tag{iii}\label{eq:fps:iii}
    \textit{own}\bigl(P,\textit{eph}(pk)\bigl) \wedge\ Q\ \textit{says}\ \bigl\{P\xleftrightarrow{K}Q\bigl\}_{pk} \rightarrow \\
    P\ \textit{allows}\ \textit{fps}\bigl(P\xleftrightarrow{K}Q\bigl) 
\end{multline*}
If the condition \eqref{eq:fps:iii} holds, then the key \(P\xleftrightarrow{F_{0}(\widebreve{PK})}Q\) satisfies Perfect Forward Secrecy. Note that in this case, the public key $pk$ is a ciphering public key. 
\end{axiom*}

\subsection{Formal Verification of 5G-AKA-HPQC using SVO Logic}
In this section, we perform a comprehensive security analysis of the 5G-AKA-HPQC protocol through formal verification, leveraging an enhanced SVO logic. This approach is augmented by including a novel perfect forward secrecy axiom formulated to examine and rigorously validate long-term security properties within the protocol.

\subsubsection{Initial State Assumptions}
As the first step, we define the initial state assumptions regarding the 5G-AKA-HPQC protocol. 
\begingroup\leqnos
\begin{equation} \tag{A11}\label{eq:A11} 
    \textit{HN believes}\ PK_{\psi}(\HN,pk_{\HN}^{\psi}) 
\end{equation}
\begin{equation} \tag{A12}\label{eq:A12} 
    \textit{HN believes}\ PK_{\delta}(\HN,pk_{\HN}^{\delta}) 
\end{equation}
\begin{equation} \tag{A13}\label{eq:A13} 
    \textit{HN believes UE controls}\ PK_{\psi}(\HN,pk1_{\UE}) 
\end{equation}
\begin{equation} \tag{A14}\label{eq:A14} 
    \textit{HN believes}\ PK_{\delta}(\HN,pk_{\HN}) 
\end{equation}
\begin{equation} \tag{A15}\label{eq:A15}
    \textit{HN believes UE} \xleftrightarrow{ss1_{\HN}}\HN 
\end{equation}
\begin{equation} \tag{H1}\label{eq:H1}
    \textit{HN believes } PK_{\delta}(\UE,{\langle pk2_{\UE} \rangle}_{*\HN}) 
\end{equation}
\begin{equation} \tag{H2}\label{eq:H2}
    \textit{HN believes UE}\xleftrightarrow{{\langle {ss1}_{\UE} \rangle}_{*\HN}} \HN 
\end{equation}
\begin{equation} \tag{H3}\label{eq:H3}
    \textit{HN believes } \textit{fresh}(pk2_{\UE}) 
\end{equation}
\begin{equation} \tag{A21}\label{eq:A21}
    \textit{SN believes SN}\xleftrightarrow{Ksh} \HN 
\end{equation}
\begin{equation} \tag{A22}\label{eq:A22}
    \textit{SN believes fresh}(n)
\end{equation}
\begin{equation} \tag{A23}\label{eq:A23} 
    \textit{SN believes HN controls UE}\xleftrightarrow{\textit{HXRES}^{*}}\SN
\end{equation}
\begin{equation} \tag{A24}\label{eq:A24} 
    \textit{SN believes HN controls fresh}(\textit{HXRES}^{*})
\end{equation}
\begin{equation} \tag{A31}\label{eq:A31}
    \textit{UE believes UE}\xleftrightarrow{K} \HN 
\end{equation}
\begin{equation} \tag{A32}\label{eq:A32}
    \textit{UE believes }\textit{fresh}(\textit{SQN}) 
\end{equation}
\begin{equation} \tag{A33}\label{eq:A33} 
    \textit{UE believes HN controls}\ PK_{\delta}(\HN,pk_{\HN}) 
\end{equation}
\begin{equation} \tag{A34}\label{eq:A34} 
    \textit{UE believes }PK_{\delta}(\UE,pk2_{\UE}) 
\end{equation}
\begin{equation} \tag{A35}\label{eq:A35} 
    \textit{UE believes }PK_{\psi}(\UE,pk1_{\UE}) 
\end{equation}
\begin{equation} \tag{A36}\label{eq:A36} 
    \textit{UE believes HN controls UE}\xleftrightarrow{ss1_{\HN}}\HN
\end{equation}
\begin{equation} \tag{A37}\label{eq:A37} 
    \textit{UE believes eph}(\{ pk1_{\UE},pk2_{\UE} \})
\end{equation}
\begin{equation} \tag{A51}\label{eq:A51}
    \textit{HN believes UE}\xleftrightarrow{K} \HN 
\end{equation}
\begin{equation} \tag{A52}\label{eq:A52}
    \textit{HN believes fresh}(pk_{\HN}) 
\end{equation}
\begin{equation} \tag{A53}\label{eq:A53} 
    \textit{HN believes UE controls}\ PK_{\delta}(\UE,pk2_{\UE}) 
\end{equation}
\begin{equation} \tag{A54}\label{eq:A54} 
    \textit{HN believes UE controls UE}\xleftrightarrow{ss1_{\UE}}\HN
\end{equation}
\begin{equation} \tag{A55}\label{eq:A55} 
    \textit{HN believes UE controls fresh}(pk2_{\UE})
\end{equation}
\begin{equation} \tag{A56}\label{eq:A56} 
    \textit{HN believes eph}(pk_{\HN})
\end{equation}
\begin{equation} \tag{A61}\label{eq:A61} 
    \textit{SN believes fresh}(m)
\end{equation}
\begin{equation} \tag{A62}\label{eq:A62} 
    \textit{SN believes HN controls UE}\xleftrightarrow{K_{\textit{SEAF}}}\HN
\end{equation}
\endgroup

\subsubsection{Received Message Assumption} 
With the initial assumptions established, the 5G-AKA-HPQC protocol is subsequently annotated as follows
\begingroup\leqnos
\begin{equation} \tag{R1} \label{eq:R1}
    \begin{split}
    \textit{HN received }\left(
    \begin{gathered}
    {\{ss1_{\UE}\}}_{PK_{\HN}^{\psi}}, pk2_{\UE}, \\
    {\{ \textit{SUPI}, pk1_{\UE} \}}_{k_1}, \\
    {\Bigl\{ {\{  \textit{SUPI},pk1_{\UE} \} }_{k_1} \Bigl\}}_{k_2}, ID_{\SN}
    \end{gathered}
    \right)
    \end{split}
\end{equation}
\begin{equation} \tag{R2} \label{eq:R2}
    \begin{split}
        \textit{SN received } {\left\{
        \begin{gathered}
            \textit{RAND},({\{\textit{SQN}\}}_{AK},\textit{AMF},\textit{MAC}), \\
            \textit{HXRES}^*,n
        \end{gathered}
        \right\}}_{Ksh}
    \end{split}
\end{equation}
\begin{equation} \tag{R3} \label{eq:R3}
    \textit{UE received }\bigl(\textit{RAND}, ({\{\textit{SQN}\}}_{AK}, \textit{AMF},\textit{MAC} )\bigl)
\end{equation}
\begin{equation} \tag{R4} \label{eq:R4}
    \textit{SN received RES}^*
\end{equation}
\begin{equation} \tag{R5} \label{eq:R5}
    \textit{HN received RES}^*
\end{equation}
\begin{equation} \tag{R6} \label{eq:R6}
    \textit{SN received }{\{ \textit{SUPI},K_{\textit{SEAF},m} \}}_{Ksh}
\end{equation}
\endgroup
\subsubsection{Comprehension Assumptions}
This step outlines how each principal comprehends the received message, which is detailed below.
\begingroup\leqnos
\begin{multline*} \tag{C1} \label{eq:C1}
    \textit{HN believes HN received }  \\ \left(
    \begin{gathered}
    {\{ { \langle ss1_{\UE} \rangle}_{*\HN} \}}_{PK_{\HN}^{\psi}},  
    {\langle pk2_{\UE} \rangle}_{*\HN}, \\
    {\{ \textit{SUPI}, {\langle pk1_{\UE} \rangle}_{*\HN} \}}_{ {\langle k_1 \rangle}_{*\HN} }, \\
    {\Bigl\{ {\{  \textit{SUPI}, {\langle pk1_{\UE} \rangle}_{*\HN} \} }_{{\langle k_1 \rangle}_{*\HN}} \Bigl\}}_{{\langle k_2 \rangle}_{*\HN}}, ID_{\SN}
    \end{gathered}
    \right)
\end{multline*}
\begin{multline*} \tag{C2} \label{eq:C2}
    \textit{SN believes SN received }  \\ {\left\{
    \begin{gathered}
        {\langle \textit{RAND},({\{\textit{SQN}\}}_{AK},\textit{AMF},\textit{MAC}) \rangle}_{*\SN}, \\
        {\langle \textit{HXRES}^* \rangle}_{*\SN},n
    \end{gathered}
    \right\}}_{Ksh}
\end{multline*}
\begin{multline*} \tag{C3} \label{eq:C3}
    \textit{UE believes UE received }  \\ \left(
    \textit{RAND}_{\textit{2UE}} \left(
    \begin{gathered}
        {\{\textit{SQN}\}}_{{\langle AK \rangle}_{*\UE}}, \textit{AMF},\\
        \left\{ 
        \begin{gathered}
            \textit{SQN}, \textit{AMF},\\
            \textit{RAND}_{\textit{2UE}}, {\langle \textit{HPK} \rangle}_{*\UE}
        \end{gathered}
        \right\}_{K}
    \end{gathered}
    \right)
    \right)
\end{multline*}
\begin{equation} \tag{C4} \label{eq:C4}
    \textit{SN believes SN received }{\langle \textit{RES}^* \rangle}_{*\SN}
\end{equation}
\begin{multline*} \tag{C5} \label{eq:C5}
    \textit{HN believes HN received }  \\ {\left\{
    \begin{gathered}
        ID_{\SN},\textit{RAND}_{\textit{2HN}}, 
        {\{ \textit{RAND}_{\textit{2HN}},\textit{HPK}\}}_{K}\ \textit{from UE}
    \end{gathered}
    \right\}}_{ICK}
\end{multline*}
\begin{multline*} \tag{C6} \label{eq:C6}
    \textit{SN believes SN received } \\ 
    { \Big\{ \{  {\langle \textit{SUPI} \rangle}_{*\SN}, { \langle K_{\textit{SEAF}} \rangle }_{*\SN},m  \} \Big\} }_{Ksh}
\end{multline*}
\endgroup
Where: 
\begin{equation} \notag
\begin{split}
    \textit{RAND} &= ( {\{ ss1_{\HN} \}}_{pk1_{\UE}}, pk_{\HN} )  \\
    \textit{MAC} &= {\{ \textit{SQN},\textit{AMF},\textit{RAND},\textit{HPK} \}}_{K} \\
    \textit{RES}^* &= {\{ ID_{\SN},\textit{RAND}, {\{ \textit{RAND},\textit{HPK}\}}_{K}\ \textit{from UE} \}}_{ICK} \\
    \textit{RAND}_{\textit{2UE}} &= \Big( {\{ {\langle ss1_{\HN} \rangle}_{*\UE} \}}_{pk1_{\UE}}, {\langle pk_{\HN} \rangle}_{*\UE} \Big) \\
    \textit{RAND}_{\textit{2HN}} &= \Big( {\{ ss1_{\HN} \}}_{ {\langle pk1_{\UE} \rangle}_{*\HN} }, pk_{\HN} \Big)  \\
\end{split}
\end{equation}

\subsubsection{Interpretation Assumptions}
In this step, we illustrate how each principal interprets the comprehended message, as described below.
\begingroup\leqnos
\begin{multline*} \tag{I1} \label{eq:I1}
    \textit{HN believes HN received }  \\ 
    \left(
    \begin{gathered}
    {\{ { \langle ss1_{\UE} \rangle}_{*\HN} \}}_{PK_{\HN}^{\psi}},  
    {\langle pk2_{\UE} \rangle}_{*\HN}, \\
    {\{ \textit{SUPI}, {\langle pk1_{\UE} \rangle}_{*\HN} \}}_{ {\langle k_1 \rangle}_{*\HN} }, \\
    {\Bigl\{ {\{  \textit{SUPI}, {\langle pk1_{\UE} \rangle}_{*\HN} \} }_{{\langle k_1 \rangle}_{*\HN}} \Bigl\}}_{{\langle k_2 \rangle}_{*\HN}}, ID_{\SN}
    \end{gathered}
    \right) \\ \rightarrow 
    \textit{HN believes HN received } \hphantom{spaces}\hphantom{spaces}\hphantom{spaces}\hphantom{spaces}\\ 
    \left(
    \begin{gathered}
    {\Big\{ \UE\xleftrightarrow{ {\langle ss1_{\UE} \rangle}_{*\HN} }\HN \Big\}}_{PK_{\HN}^{\psi}}, \\
    PK_{\delta}(\UE,{\langle pk2_{\UE} \rangle}_{*\HN}), \\
    {\left\{ 
    \begin{gathered}
        {\left\{ 
        \begin{gathered}
        \textit{SUPI},PK_{\psi}(\UE, {\langle {pk1}_{\UE} \rangle}_{*\HN} ), \\
        PK_{\delta}(\UE, {\langle {pk2}_{\UE} \rangle}_{*\HN} )
        \end{gathered}
        \right\}}_{{\langle k_1 \rangle}_{*\HN}}, \\
        PK_{\delta}(\UE, {\langle {pk2}_{\UE} \rangle}_{*\HN} )
    \end{gathered}
    \right\}}_{{\langle k_2 \rangle}_{*\HN} }, \\
    ID_{\SN}
    \end{gathered}
    \right)
\end{multline*}
\begin{multline*} \tag{I2} \label{eq:I2}
    \textit{SN believes SN received }  \\ {\left\{
    \begin{gathered}
        {\langle \textit{RAND},({\{\textit{SQN}\}}_{AK},\textit{AMF},\textit{MAC}) \rangle}_{*\SN}, \\
        {\langle \textit{HXRES}^* \rangle}_{*\SN},n
    \end{gathered}
    \right\}}_{Ksh} \\ \rightarrow 
    \textit{SN believes SN received } \hphantom{spaces}\hphantom{spaces}\hphantom{spaces}\hphantom{spaces}\\
    {\left\{
    \begin{gathered}
        {\langle \textit{RAND},({\{\textit{SQN}\}}_{AK},\textit{AMF},\textit{MAC}) \rangle}_{*\SN}, \\
        \UE\xleftrightarrow{{\langle \textit{HXRES}^* \rangle}_{*\SN}}\SN, \\
        \textit{fresh}({\langle \textit{HXRES}^* \rangle}_{*\SN}),n
    \end{gathered}
    \right\}}_{Ksh}
\end{multline*}

\begin{multline*} \tag{I3} \label{eq:I3}
    \textit{UE believes UE received }  \\ \left(
    \textit{RAND}_{\textit{2UE}}, \left(
    \begin{gathered}
        {\{\textit{SQN}\}}_{{\langle AK \rangle}_{*\UE}}, \textit{AMF},\\
        \left\{ 
        \begin{gathered}
            \textit{SQN}, \textit{AMF},\\
            \textit{RAND}_{\textit{2UE}}, {\langle \textit{HPK} \rangle}_{*\UE}
        \end{gathered}
        \right\}_{K}
    \end{gathered}
    \right)
    \right) \\ \rightarrow 
    \textit{UE believes UE received } \hphantom{spaces}\hphantom{spaces}\hphantom{spaces}\\
    {\left\{
    \begin{gathered}
        \textit{SQN}, \textit{AMF}, {\Big\{ \UE\xleftrightarrow{ {\langle ss1_{\HN} \rangle }_{*\UE} }\HN \Big\}}_{pk1_{\UE}},\\
        PK_{\delta}(\HN, {\langle {pk}_{\HN} \rangle}_{*\UE}),\textit{eph}({\langle {pk}_{\HN} \rangle}_{*\UE}),\\
        \UE\xleftrightarrow{{\langle \textit{HPK} \rangle}_{*\UE}}\HN
    \end{gathered}
    \right\}}_{K}
\end{multline*}
The first and second values of \(\textit{RAND}_{\textit{2UE}}\) and \({\{\textit{SQN}\}}_{{\langle AK \rangle}_{*\UE}}\) is omitted because they overlap with the content of last value.
\begin{multline*} \tag{I4} \label{eq:I4}
    (\textit{SN believes SN received }{\langle \textit{RES}^* \rangle}_{*\SN})\ \\ \wedge 
    (\textit{SN believes UE}\xleftrightarrow{{\langle \textit{HXRES}^* \rangle}_{*\SN}}\SN ) \\
    \rightarrow \textit{SN believes SN received} \hphantom{spaces}\hphantom{spaces}\hphantom{spaces}\\
    {\{ {\langle \textit{RES}^* \rangle }_{*\SN}, {\langle \textit{HXRES}^* \rangle}_{*\SN} \}}_{{\langle \textit{HXRES}^* \rangle}_{*\SN}}
\end{multline*}
\begin{multline*} \tag{I5} \label{eq:I5}
    \textit{HN believes HN received }  \\ {\big\{
        ID_{\SN},\textit{RAND}_{\textit{2HN}}, 
        {\{ \textit{RAND}_{\textit{2HN}},\textit{HPK}\}}_{K}\ \textit{from UE} 
    \big\}}_{\textit{ICK}}\ \\ \wedge 
    \Big(\textit{HN believes UE}\xleftrightarrow{\textit{ICK}}\HN \Big) \\
    \rightarrow \textit{HN believes HN received} \hphantom{spaces}\hphantom{spaces}\hphantom{spaces}\hphantom{spaces}\\
    {\left\{
    \begin{gathered}
        ID_{\SN}, PK_{\psi}(\HN,pk_{\HN}), \\
        {\left\{
        \begin{gathered}
            { \Big\{\UE\xleftrightarrow{ss1_{\HN}}\HN \Big\}}_{{\langle pk1_{\UE} \rangle}_{*\HN}}, PK_{\delta}(\HN,pk_{\HN}),\\
            PK_{\psi}(\UE, {\langle {pk1}_{\UE} \rangle}_{*\HN} ), \textit{ fresh}( {\langle {pk1}_{\UE} \rangle}_{*\HN}), \\
            PK_{\delta}(\UE, {\langle {pk2}_{\UE} \rangle}_{*\HN}), \\
            \textit{eph}(\{ {\langle {pk1}_{\UE} \rangle}_{*\HN},{\langle {pk2}_{\UE} \rangle}_{*\HN} \}),\\
            \UE\xleftrightarrow{ {\langle ss1_{\UE} \rangle}_{*\HN} }\HN, \UE\xleftrightarrow{\textit{HPK}}\HN
        \end{gathered}
        \right\}}_{\textit{K}}
    \end{gathered}
    \right\}}_{\textit{ICK}}
\end{multline*}

In the \(\textit{RES}^*\) message, \(\textit{RAND}_{\textit{2HN}}\) is omitted because it overlaps with the same value in the included  \(\textit{RES}\), leaving only \(pk_{\HN}\) that can verify freshness. The \(\textit{RES}\) value explicitly contains the \(\textit{RAND}\) value and the \(\textit{HPK}\), as well as with the UE's trust in the public keys \(pk1_{\UE}\) and \(pk2_{\UE}\), the freshness of \(pk1_{\UE}\), and the secret key generation key material \(ss1_{\UE}\) used SUCI encryption. These elements are inherently linked to the SUCI transmission process, necessitating their inclusion. Consequently, if \(\textit{RES}^*\) and \(\textit{RES}\) are validated, it also confirms the freshness and validity of \(pk1_{\UE}\), \(pk2_{\UE}\) and \(ss1_{\UE}\). Finally, since the UE includes the belief that \(pk1_{\UE}\) and \(pk2_{\UE}\) are ephemeral public keys in the \(\textit{RES}^*\) message, \( \textit{eph}(\{ {\langle {pk1}_{\UE} \rangle}_{*\HN},{\langle {pk2}_{\UE} \rangle}_{*\HN} \})\) is added.

\begin{multline*} \tag{I6} \label{eq:I6}
    \textit{SN believes SN received } \\ 
    { \Big\{ \{  {\langle \textit{SUPI} \rangle}_{*\SN}, { \langle K_{\textit{SEAF}} \rangle }_{*\SN},m  \} \Big\} }_{Ksh} \\
    \rightarrow \textit{SN believes SN received } \hphantom{spaces}\hphantom{spaces}\hphantom{spaces}\hphantom{spaces}\\ 
    { \Big\{   {\langle \textit{SUPI} \rangle}_{*\SN}, \UE\xleftrightarrow{ { \langle K_{\textit{SEAF}} \rangle }_{*\SN} }\SN   ,m  \Big\} }_{Ksh}
\end{multline*}

Note that: 
\begin{equation} \notag
\begin{split}
    \textit{eph}(\{ pk1_{\UE},pk2_{\UE} \}) &=  \textit{own}(\UE,\{ pk1_{\UE},pk2_{\UE},pk_{\HN} \}) \\
    \textit{eph}(pk_{\HN}) &= \textit{own}(\HN,\{ pk1_{\UE},pk2_{\UE},pk_{\HN} \}) \\
    \textit{RAND}_{\textit{2UE}} &= ( {\{ {\langle ss1_{\HN} \rangle}_{*\UE} \} }_{pk1_{\UE}}, {\langle pk_{\HN}\rangle }_{*\UE}) \\
    \textit{RAND}_{\textit{2HN}} &= ( {\{ ss1_{\HN} \}}_{\langle {pk1_{\UE} \rangle}_{*\HN} }, pk_{\HN} )
\end{split}
\end{equation}
\endgroup
\subsubsection{Derivation}
In the final step, we iteratively apply axioms, inference rules, and the newly incorporated forward perfect secrecy axiom until the desired results are achieved.\\
From \eqref{eq:I1}, we derive
\begingroup\leqnos
\begin{multline*} \tag{D11} \label{eq:D11}
    \textit{HN believes HN received } \\ \left( 
    \begin{gathered}
    {\Big\{ \UE\xleftrightarrow{ {\langle ss1_{\UE} \rangle}_{*\HN} }\HN \Big\}}_{PK_{\HN}^{\psi}}, \\
    PK_{\delta}(\UE,{\langle pk2_{\UE} \rangle}_{*\HN}), \\
    {\left\{ 
    \begin{gathered}
        {\left\{ 
        \begin{gathered}
        \textit{SUPI},PK_{\psi}(\UE, {\langle {pk1}_{\UE} \rangle}_{*\HN} ), \\
        PK_{\delta}(\UE, {\langle {pk2}_{\UE} \rangle}_{*\HN} )
        \end{gathered}
        \right\}}_{{\langle k_1 \rangle}_{*\HN}}, \\
        PK_{\delta}(\UE, {\langle {pk2}_{\UE} \rangle}_{*\HN} )
    \end{gathered}
    \right\}}_{{\langle k_2 \rangle}_{*\HN} }, \\
    ID_{\SN}
    \end{gathered}
    \right)
\end{multline*}
\begin{flushright}
    by \eqref{eq:C1}, \eqref{eq:I1}, MP
\end{flushright}
\begin{equation} \tag{D12} \label{eq:D12}
    \textit{HN believes HN has UE}\xleftrightarrow{ {\langle ss1_{\UE} \rangle}_{*\HN} }\HN
\end{equation}
\begin{flushright}
    by \eqref{eq:D11}, \eqref{eq:A11}, RA, PA, BA
\end{flushright}
\begin{equation} \tag{D13} \label{eq:D13}
    \textit{HN believes HN has }PK_{\delta}(\UE, { \langle pk2_{\UE} \rangle }_{*\HN} )
\end{equation}
\begin{flushright}
    by \eqref{eq:D11}, RA, PA, BA
\end{flushright}
\begin{equation} \tag{D14} \label{eq:D14}
    \textit{HN believes UE}\xleftrightarrow{pk_{\HN}^{-1}\cdot {\langle pk2_{\UE} \rangle}_{*\HN} \cdot G}\HN
\end{equation}
\begin{flushright}
    by \eqref{eq:A12}, \eqref{eq:H1}, KA, BA
\end{flushright}
\begin{equation} \tag{D15} \label{eq:D15}
    \textit{HN believes UE}\xleftrightarrow{k_1}\HN
\end{equation}
\begin{flushright}
    by \eqref{eq:D14}, \eqref{eq:H2}
\end{flushright}
\begin{equation} \tag{D16} \label{eq:D16}
    \textit{HN believes UE}\xleftrightarrow{k_2}\HN
\end{equation}
\begin{flushright}
    by \eqref{eq:D14}, \eqref{eq:H2}
\end{flushright}
\begin{multline*} \tag{D17} \label{eq:D17}
    \textit{HN believes UE said } \\ \left( 
    \begin{gathered}
        {\left\{ 
        \begin{gathered}
        \textit{SUPI},PK_{\psi}(\UE, {\langle {pk1}_{\UE} \rangle}_{*\HN} ), \\
        PK_{\delta}(\UE, {\langle {pk2}_{\UE} \rangle}_{*\HN} )
        \end{gathered}
        \right\}}_{{\langle k_1 \rangle}_{*\HN}}, \\
        PK_{\delta}(\UE, {\langle {pk2}_{\UE} \rangle}_{*\HN} )
    \end{gathered}
    \right)
\end{multline*}
\begin{flushright}
    by \eqref{eq:D11}, RA, \eqref{eq:D16}, SAA, BA
\end{flushright}
\begin{multline*} \tag{D18} \label{eq:D18}
    \textit{HN believes UE says } \\ \left( 
    \begin{gathered}
        {\left\{ 
        \begin{gathered}
        \textit{SUPI},PK_{\psi}(\UE, {\langle {pk1}_{\UE} \rangle}_{*\HN} ), \\
        PK_{\delta}(\UE, {\langle {pk2}_{\UE} \rangle}_{*\HN} )
        \end{gathered}
        \right\}}_{k_1}, \\
        PK_{\delta}(\UE, {\langle {epk}_{\UE} \rangle}_{*\HN} )
    \end{gathered}
    \right)
\end{multline*}
\begin{flushright}
    by \eqref{eq:D17}, \eqref{eq:H3}, FR, NV, BA
\end{flushright}
\begin{multline*} \tag{D19} \label{eq:D19}
    \textit{HN believes HN received } \\ 
        {\left\{ 
        \begin{gathered}
        \textit{SUPI},PK_{\psi}(\UE, {\langle {pk1}_{\UE} \rangle}_{*\HN} ), \\
        PK_{\delta}(\UE, {\langle {pk2}_{\UE} \rangle}_{*\HN} )
        \end{gathered}
        \right\}}_{k_1}
\end{multline*}
\begin{flushright}
    by \eqref{eq:D17}, SA, BA
\end{flushright}
\begin{multline*} \tag{D1a} \label{eq:D1a}
    \textit{HN believes UE says } \\ 
        \left( 
        \begin{gathered}
        \textit{SUPI},PK_{\psi}(\UE, {\langle {pk1}_{\UE} \rangle}_{*\HN} ), \\
        PK_{\delta}(\UE, {\langle {pk2}_{\UE} \rangle}_{*\HN} )
        \end{gathered}
        \right)
\end{multline*}
\begin{flushright}
    by \eqref{eq:D19}, \eqref{eq:D15}, SAA, \eqref{eq:H3}, FR, NV, BA
\end{flushright}
\begin{equation} \tag{D1b} \label{eq:D1b}
    \textit{HN believes UE says SUPI}
\end{equation}
\begin{flushright}
    by \eqref{eq:D1a}, SA, BA
\end{flushright}
\begin{equation} \tag{D1c} \label{eq:D1c}
    \textit{HN believes }PK_{\psi}(\UE, {\langle {pk1}_{\UE} \rangle}_{*\HN} )
\end{equation}
\begin{flushright}
    by \eqref{eq:D1a}, SA, \eqref{eq:A13}, JR, BA
\end{flushright}
\begin{equation} \tag{D1d} \label{eq:D1d}
    \textit{HN believes UE}\xleftrightarrow{PK_{\HN}^{-1}\cdot {\langle pk2_{\UE} \rangle}_{*\HN} \cdot G}\HN
\end{equation}
\begin{flushright}
    by \eqref{eq:A14}, \eqref{eq:H1}, KA, BA
\end{flushright}
\begin{equation} \tag{D1e} \label{eq:D1e}
    \textit{HN believes UE}\xleftrightarrow{\textit{HPK}}\HN
\end{equation}
\begin{flushright}
    by \eqref{eq:A15}, \eqref{eq:D1d}
\end{flushright}
Where: 
\begin{equation} \notag
\begin{split}
    \textit{HPK} &= H(  ss1_{\HN}, pk_{\HN}^{-1} \cdot pk2_{\UE} \cdot G  )  \\
\end{split}
\end{equation}
\endgroup

The derivation \eqref{eq:D13} and \eqref{eq:D1c} cannot be extended further and concluded at this point. As a result, \eqref{eq:H1} is introduced to continue the analysis. Similarly, \eqref{eq:D17} and \eqref{eq:D19} cannot be evolved further and must end here. Therefore, \eqref{eq:H2} and \eqref{eq:H3} are added to advance the analysis. \eqref{eq:H1}, \eqref{eq:H2} and \eqref{eq:H3} can be vulnerabilities in this protocol.

From \eqref{eq:I2}, we derive
\begingroup\leqnos
\begin{multline*} \tag{D21} \label{eq:D21}
    \textit{SN believed SN received } \\
    {\left\{
    \begin{gathered}
        {\langle \textit{RAND},({\{\textit{SQN}\}}_{AK},\textit{AMF},\textit{MAC}) \rangle}_{*\SN}, \\
        \UE\xleftrightarrow{{\langle \textit{HXRES}^* \rangle}_{*\SN}}\SN, \\
        \textit{fresh}({\langle \textit{HXRES}^* \rangle}_{*\SN}),n
    \end{gathered}
    \right\}}_{Ksh}
\end{multline*}
\begin{flushright}
    by \eqref{eq:C2}, \eqref{eq:I2}, MP
\end{flushright}
\begin{multline*} \tag{D22} \label{eq:D22}
    \textit{SN believed HN said }\\
    \left(
    \begin{gathered}
        {\langle \textit{RAND},({\{\textit{SQN}\}}_{AK},\textit{AMF},\textit{MAC}) \rangle}_{*\SN}, \\
        \UE\xleftrightarrow{{\langle \textit{HXRES}^* \rangle}_{*\SN}}\SN, \\
        \textit{fresh}({\langle \textit{HXRES}^* \rangle}_{*\SN}),n
    \end{gathered}
    \right)
\end{multline*}
\begin{flushright}
    by \eqref{eq:D21}, \eqref{eq:A21}, SAA, BA
\end{flushright}
\begin{multline*} \tag{D23} \label{eq:D23}
    \textit{SN believed HN says }\\
    \left(
    \begin{gathered}
        {\langle \textit{RAND},({\{\textit{SQN}\}}_{AK},\textit{AMF},\textit{MAC}) \rangle}_{*\SN}, \\
        \UE\xleftrightarrow{{\langle \textit{HXRES}^* \rangle}_{*\SN}}\SN, \\
        \textit{fresh}({\langle \textit{HXRES}^* \rangle}_{*\SN}),n
    \end{gathered}
    \right)
\end{multline*}
\begin{flushright}
    by \eqref{eq:D22}, \eqref{eq:A22}, FR, NV, BA
\end{flushright}
\begin{equation} \tag{D24} \label{eq:D24}
    \textit{SN believes HN says UE}\xleftrightarrow{ {\langle \textit{HXRES}^* \rangle}_{*\SN} }\SN
\end{equation}
\begin{flushright}
    by \eqref{eq:D23}, SA, BA
\end{flushright}
\begin{equation} \tag{D25} \label{eq:D25}
    \textit{SN believes UE}\xleftrightarrow{ {\langle \textit{HXRES}^* \rangle}_{*\SN} }\SN
\end{equation}
\begin{flushright}
    by \eqref{eq:D24}, \eqref{eq:A23}, JR, BA
\end{flushright}
\begin{equation} \tag{D26} \label{eq:D26}
    \textit{SN believes fresh}( {\langle \textit{HXRES}^* \rangle}_{*\SN} )
\end{equation}
\begin{flushright}
    by \eqref{eq:D23}, SA, \eqref{eq:A24}, JR, BA
\end{flushright}
\endgroup

From \eqref{eq:I3}, we derive
\begingroup\leqnos
\begin{multline*} \tag{D31} \label{eq:D31}
    \textit{UE believes UE received } \\
    {\left\{
    \begin{gathered}
        \textit{SQN}, \textit{AMF}, {\{ \UE\xleftrightarrow{ {\langle ss1_{\HN} \rangle }_{*\UE} }\HN \}}_{pk1_{\UE}},\\
        PK_{\delta}(\HN, {\langle {pk}_{\HN} \rangle}_{*\UE}),\textit{eph}({\langle {pk}_{\HN} \rangle}_{*\UE}),\\
        \UE\xleftrightarrow{{\langle \textit{HPK} \rangle}_{*\UE}}\HN
    \end{gathered}
    \right\}}_{K}
\end{multline*}
\begin{flushright}
    by \eqref{eq:C3}, \eqref{eq:I3}, MP
\end{flushright}
\begin{multline*} \tag{D32} \label{eq:D32}
    \textit{UE believes UE said } \\
    \left(
    \begin{gathered}
        \textit{SQN}, \textit{AMF}, {\{ \UE\xleftrightarrow{ {\langle ss1_{\HN} \rangle }_{*\UE} }\HN \}}_{pk1_{\UE}},\\
        PK_{\delta}(\HN, {\langle {pk}_{\HN} \rangle}_{*\UE}),\textit{eph}({\langle {pk}_{\HN} \rangle}_{*\UE}),\\
        \UE\xleftrightarrow{{\langle \textit{HPK} \rangle}_{*\UE}}\HN
    \end{gathered}
    \right)
\end{multline*}
\begin{flushright}
    by \eqref{eq:D31}, \eqref{eq:A31}, SAA, BA
\end{flushright}
\begin{multline*} \tag{D33} \label{eq:D33}
    \textit{UE believes HN says } \\
    \left(
    \begin{gathered}
        \textit{SQN}, \textit{AMF}, {\{ \UE\xleftrightarrow{ {\langle ss1_{\HN} \rangle }_{*\UE} }\HN \}}_{pk1_{\UE}},\\
        PK_{\delta}(\HN, {\langle {pk}_{\HN} \rangle}_{*\UE}),\textit{eph}({\langle {pk}_{\HN} \rangle}_{*\UE}),\\
        \UE\xleftrightarrow{{\langle \textit{HPK} \rangle}_{*\UE}}\HN
    \end{gathered}
    \right)
\end{multline*}
\begin{flushright}
    by \eqref{eq:D32}, \eqref{eq:A32}, FR, NV, BA
\end{flushright}
\begin{equation} \tag{D34} \label{eq:D34}
    \textit{UE believes }PK_{\delta}(\HN, {\langle {pk}_{\HN} \rangle}_{*\UE})
\end{equation}
\begin{flushright}
    by \eqref{eq:D33}, SA, \eqref{eq:A33}, JR, BA
\end{flushright}
\begin{equation} \tag{D35} \label{eq:D35}
    \textit{UE believes UE}\xleftrightarrow{ pk2_{\UE}^{-1} \cdot {\langle pk_{\HN} \rangle}_{*\UE} \cdot G}\HN
\end{equation}
\begin{flushright}
    by \eqref{eq:A34}, \eqref{eq:D34}, KA, BA
\end{flushright}
\begin{equation} \tag{D36} \label{eq:D36}
    \textit{UE believes HN says }{\Big\{ \UE\xleftrightarrow{ {\langle ss1_{\HN} \rangle }_{*\UE} }\HN \Big\}}_{pk1_{\UE}}
\end{equation}
\begin{flushright}
    by \eqref{eq:D33}, SA, BA
\end{flushright}
\begin{equation} \tag{D37} \label{eq:D37}
    \textit{UE believes }{ \UE\xleftrightarrow{ {\langle ss1_{\HN} \rangle }_{*\UE} }\HN }
\end{equation}
\begin{flushright}
    by \eqref{eq:D36}, \eqref{eq:A35}, RA, \eqref{eq:A36}, JR, BA
\end{flushright}
\begin{equation} \tag{D38} \label{eq:D38}
    \textit{UE believes }{\UE\xleftrightarrow{ {\langle \textit{HPK} \rangle }_{*\UE} }\HN }
\end{equation}
\begin{flushright}
    by \eqref{eq:D35}, \eqref{eq:D37}
\end{flushright}
\begin{equation} \tag{D39} \label{eq:D39}
    \textit{UE believes HN says UE}{\xleftrightarrow{ {\langle \textit{HPK} \rangle }_{*\UE} }\HN }
\end{equation}
\begin{flushright}
    by \eqref{eq:D33}, SA, BA
\end{flushright}
\begin{equation} \tag{D3a} \label{eq:D3a}
    \textit{UE believes HN says eph}({ \langle pk_{\HN} \rangle }_{*\UE})
\end{equation}
\begin{flushright}
    by \eqref{eq:D33}, SA, BA
\end{flushright}
\begin{equation} \tag{D3b} \label{eq:D3b}
    \textit{UE believes allow fps}\Big({\UE\xleftrightarrow{ {\langle \textit{HPK} \rangle }_{*\UE} }\HN }\Big)
\end{equation}
\begin{flushright}
    by \eqref{eq:A52}, \eqref{eq:D3a}, FPSA, BA
\end{flushright}
\begin{equation} \tag{D3c} \label{eq:D3c}
    \textit{UE believes }\UE\xleftrightarrow{ K_{\textit{SEAF}} }\HN 
\end{equation}
\begin{flushright}
    by \eqref{eq:A31}, \eqref{eq:D33}, SA, \eqref{eq:D38}, BA
\end{flushright}

Where: 
\begin{equation} \notag
\begin{split}
    \textit{HPK} &= H(  ss1_{\HN}, pk2_{\UE}^{-1} \cdot pk_{\HN} \cdot G  )  \\
\end{split}
\end{equation}
\endgroup

From \eqref{eq:I4}, we derive
\begingroup\leqnos
\begin{multline*} \tag{D41} \label{eq:D41}
    \textit{SN believes SN received} \\
    {\{ {\langle \textit{RES}^* \rangle }_{*\SN}, {\langle \textit{HXRES}^* \rangle}_{*\SN} \}}_{{\langle \textit{HXRES}^* \rangle}_{*\SN}}
\end{multline*}
\begin{flushright}
    by \eqref{eq:C4}, \eqref{eq:D25}, \eqref{eq:I4}, MP
\end{flushright}
\begin{equation} \tag{D42} \label{eq:D42}
    \textit{SN believes UE said } ( {\langle \textit{RES}^* \rangle }_{*\SN}, {\langle \textit{HXRES}^* \rangle}_{*\SN} )
\end{equation}
\begin{flushright}
    by \eqref{eq:D41}, \eqref{eq:D25}, SAA, BA
\end{flushright}
\begin{equation} \tag{D43} \label{eq:D43}
    \textit{SN believes UE says } ( {\langle \textit{RES}^* \rangle }_{*\SN}, {\langle \textit{HXRES}^* \rangle}_{*\SN} )
\end{equation}
\begin{flushright}
    by \eqref{eq:D42}, \eqref{eq:D26}, FR, NV, BA
\end{flushright}
\begin{equation} \tag{D44} \label{eq:D44}
    \textit{SN believes UE says }  {\langle \textit{RES}^* \rangle }_{*\SN}
\end{equation}
\begin{flushright}
    by \eqref{eq:D43}, SA, BA
\end{flushright}
\endgroup

From \eqref{eq:I5}, we derive
\begingroup\leqnos
\begin{equation} \tag{D51} \label{eq:D51}
    \textit{HN believes UE}{\xleftrightarrow{ \textit{ICK} }\HN }
\end{equation}
\begin{flushright}
    by \eqref{eq:D1e}, \eqref{eq:A14}, \eqref{eq:A15}, \eqref{eq:D1e}, \eqref{eq:A51}
\end{flushright}
\begin{multline*} \tag{D52} \label{eq:D52}
    \textit{HN believes HN received} \\
    {\left\{
    \begin{gathered}
        ID_{\SN}, PK_{\psi}(\HN,pk_{\HN}), \\
        {\left\{
        \begin{gathered}
            { \{\UE\xleftrightarrow{ss1_{\HN}}\HN \}}_{{\langle pk1_{\UE} \rangle}_{*\HN}}, PK_{\delta}(\HN,pk_{\HN}),\\
            PK_{\psi}(\UE, {\langle {pk1}_{\UE} \rangle}_{*\HN} ), \textit{fresh}( {\langle {pk2}_{\UE} \rangle}_{*\HN}), \\
            PK_{\delta}(\UE, {\langle {pk2}_{\UE} \rangle}_{*\HN}), \\
            \textit{eph}(\{ {\langle {pk1}_{\UE} \rangle}_{*\HN},{\langle {pk2}_{\UE} \rangle}_{*\HN} \}),\\
            \UE\xleftrightarrow{ {\langle ss1_{\UE} \rangle}_{*\HN} }\HN, \UE\xleftrightarrow{\textit{HPK}}\HN
        \end{gathered}
        \right\}}_{\textit{K}}
    \end{gathered}
    \right\}}_{\textit{ICK}}
\end{multline*}
\begin{flushright}
    by \eqref{eq:C5}, \eqref{eq:D51}, \eqref{eq:I5}, MP
\end{flushright}
\begin{multline*} \tag{D53} \label{eq:D53}
    \textit{HN believes UE says} \\
    \left(
    \begin{gathered}
        ID_{\SN}, PK_{\psi}(\HN,pk_{\HN}), \\
        {\left\{
        \begin{gathered}
            { \{\UE\xleftrightarrow{ss1_{\HN}}\HN \}}_{{\langle pk1_{\UE} \rangle}_{*\HN}}, PK_{\delta}(\HN,pk_{\HN}),\\
            PK_{\psi}(\UE, {\langle {pk1}_{\UE} \rangle}_{*\HN} ), \textit{fresh}( {\langle {pk2}_{\UE} \rangle}_{*\HN}), \\
            PK_{\delta}(\UE, {\langle {pk2}_{\UE} \rangle}_{*\HN}), \\
            \textit{eph}(\{ {\langle {pk1}_{\UE} \rangle}_{*\HN},{\langle {pk2}_{\UE} \rangle}_{*\HN} \}),\\
            \UE\xleftrightarrow{ {\langle ss1_{\UE} \rangle}_{*\HN} }\HN, \UE\xleftrightarrow{\textit{HPK}}\HN
        \end{gathered}
        \right\}}_{\textit{K}}
    \end{gathered}
    \right)
\end{multline*}
\begin{flushright}
    by \eqref{eq:D51}, \eqref{eq:D52}, SAA, \eqref{eq:A52}, FR, NV, BA
\end{flushright}
\begin{multline*} \tag{D54} \label{eq:D54}
    \textit{HN believes UE says} \\
        \left(
        \begin{gathered}
            { \{\UE\xleftrightarrow{ss1_{\HN}}\HN \}}_{{\langle pk1_{\UE} \rangle}_{*\HN}}, PK_{\delta}(\HN,pk_{\HN}),\\
            PK_{\psi}(\UE, {\langle {pk1}_{\UE} \rangle}_{*\HN} ), \textit{fresh}( {\langle {pk2}_{\UE} \rangle}_{*\HN}), \\
            PK_{\delta}(\UE, {\langle {pk2}_{\UE} \rangle}_{*\HN}), \\
            \textit{eph}(\{ {\langle {pk1}_{\UE} \rangle}_{*\HN},{\langle {pk2}_{\UE} \rangle}_{*\HN} \}),\\
            \UE\xleftrightarrow{ {\langle ss1_{\UE} \rangle}_{*\HN} }\HN, \UE\xleftrightarrow{\textit{HPK}}\HN
        \end{gathered}
        \right)
\end{multline*}
\begin{flushright}
    by \eqref{eq:D53}, SA, \eqref{eq:D51}, SAA, \eqref{eq:A52}, FR, NV, BA
\end{flushright}
\begin{equation}  \tag{D55} \label{eq:D55}
    \textit{HN believes HN says UE}\xleftrightarrow{\textit{HPK}}\HN
\end{equation}
\begin{flushright}
    by \eqref{eq:D54}, SA, BA
\end{flushright}
\begin{equation}  \tag{D56} \label{eq:D56}
    \textit{HN believes }PK_{\delta}(\UE, {\langle {pk2}_{\UE} \rangle}_{*\HN})
\end{equation}
\begin{flushright}
    by \eqref{eq:D54}, SA, \eqref{eq:A53}, JR, BA
\end{flushright}
\begin{equation}  \tag{D57} \label{eq:D57}
    \textit{HN believes }\UE\xleftrightarrow{ {\langle ss1_{\UE} \rangle}_{*\HN} }\HN
\end{equation}
\begin{flushright}
    by \eqref{eq:D54}, SA, \eqref{eq:A54}, JR, BA
\end{flushright}
\begin{equation}  \tag{D58} \label{eq:D58}
    \textit{HN believes }\textit{fresh}( {\langle {pk2}_{\UE} \rangle}_{*\HN})
\end{equation}
\begin{flushright}
    by \eqref{eq:D54}, SA, \eqref{eq:A55}, JR, BA
\end{flushright}
\begin{multline*}  \tag{D59} \label{eq:D59}
    \textit{HN believes UE says }\\ 
    \textit{eph}(\{ {\langle {pk1}_{\UE} \rangle}_{*\HN},{\langle {pk2}_{\UE} \rangle}_{*\HN} \})
\end{multline*}
\begin{flushright}
    by \eqref{eq:D54}, SA, BA
\end{flushright}
\begin{equation} \tag{D5a} \label{eq:D5a}
    \textit{HN believes HN allows UE}\xleftrightarrow{\textit{HPK}}\HN
\end{equation}
\begin{flushright}
    by \eqref{eq:A55}, \eqref{eq:D55}, PFSA, BA
\end{flushright}
\begin{equation} \tag{D5b} \label{eq:D5b}
    \textit{HN believes UE says UE}\xleftrightarrow{ { \langle \textit{HPK} \rangle }_{*\HN}}\HN
\end{equation}
\begin{flushright}
    by \eqref{eq:D54}, SA, BA
\end{flushright}
\endgroup

From \eqref{eq:I6}, we derive
\begingroup\leqnos
\begin{multline*} \tag{D61} \label{eq:D61}
    \textit{SN believes SN received }\\ 
    { \Big\{   {\langle \textit{SUPI} \rangle}_{*\SN}, \UE\xleftrightarrow{ { \langle K_{\textit{SEAF}} \rangle }_{*\SN} }\SN   ,m  \Big\} }_{Ksh}
\end{multline*}
\begin{flushright}
    by \eqref{eq:C6}, \eqref{eq:I6}, MP
\end{flushright}
\begin{multline*} \tag{D62} \label{eq:D62}
    \textit{SN believes HN said }\\ 
    \Big(   {\langle \textit{SUPI} \rangle}_{*\SN}, \UE\xleftrightarrow{ { \langle K_{\textit{SEAF}} \rangle }_{*\SN} }\SN   ,m  \Big)
\end{multline*}
\begin{flushright}
    by \eqref{eq:D61}, \eqref{eq:A21}, SAA, BA
\end{flushright}
\begin{multline*} \tag{D63} \label{eq:D63}
    \textit{SN believes HN says }\\ 
    \Big(   {\langle \textit{SUPI} \rangle}_{*\SN}, \UE\xleftrightarrow{ { \langle K_{\textit{SEAF}} \rangle }_{*\SN} }\SN   ,m  \Big)
\end{multline*}
\begin{flushright}
    by \eqref{eq:D62}, \eqref{eq:A61}, FR, NV, BA
\end{flushright}
\begin{equation} \tag{D64} \label{eq:D64}
    \textit{SN believes HN says }\UE\xleftrightarrow{ { \langle K_{\textit{SEAF}} \rangle }_{*\SN} }\SN 
\end{equation}
\begin{flushright}
    by \eqref{eq:D62}, \eqref{eq:A61}, FR, NV, BA
\end{flushright}
\begin{equation} \tag{D65} \label{eq:D65}
    \textit{SN believes }\UE\xleftrightarrow{ { \langle K_{\textit{SEAF}} \rangle }_{*\SN} }\SN 
\end{equation}
\begin{flushright}
    by \eqref{eq:D64}, \eqref{eq:A62}, JR, BA
\end{flushright}

\endgroup
\subsubsection{Result}
We derive the following lemmas from the formal analysis above, demonstrating that the security requirements are met.

\begin{lemma}
The 5G-AKA-HPQC protocol provides mutual authentication between UE and HN
\end{lemma}
\begin{proof}
The derivation \eqref{eq:D39} confirms the authentication of the HN to the UE. Similarly, the derivation \eqref{eq:D5b} establishes the authentication of the UE to the HN. Moreover, the derivation \eqref{eq:D44} shows the authentication of the SN to the UE. Consequently, this proves that mutual authentication between the UE and the 5G core networks (SN and HN) is achieved.
\end{proof}
\begin{lemma}
The 5G-AKA-HPQC protocol provides a Secure Key Exchange
\end{lemma}
\begin{proof}
Firstly, the derivations \eqref{eq:D38} and \eqref{eq:D39} confirm the UE has successfully established the belief of HPK. Conversely, the derivations \eqref{eq:D56}, \eqref{eq:D1e}, and \eqref{eq:D5b} indicate the HN has successfully obtained the belief on both the HPK and UE’s ephemeral public key. The key agreement of HPK is shown by indirect belief from the derivations \eqref{eq:D39} and \eqref{eq:D5b}, while direct belief in HPK key is provided by the derivations \eqref{eq:D38} and \eqref{eq:D1e}. Secondly, the belief \eqref{eq:D3c} shows that the UE has obtained the belief in \(K_{\textit{SEAF}}\), while the belief \eqref{eq:D65} shows the SN has similarly obtained the trust in \(K_{\textit{SEAF}}\). This proves that a secure key exchange has been successfully achieved between the UE and 5G core networks (SN and HN)
\end{proof}
\begin{lemma}
The 5G-AKA-HPQC protocol provides SUPI concealment 
\end{lemma}
\begin{proof} 
First, the hypothesis \eqref{eq:H1} assumes that the HN believes a public key \(pk2_{\UE}\). Second, the hypothesis \eqref{eq:H2} assumes that the HN believes a shared secret key \(ss1_{\HN}\) between UE and HN, and the hypothesis \eqref{eq:H3} assumes that the HN believes the freshness of \(pk2_{\UE}\). Later the derivations \eqref{eq:D56}, \eqref{eq:D57}, and \eqref{eq:D58} proves the hypothesis \eqref{eq:H1}, \eqref{eq:H2} and \eqref{eq:H3} respectively. This implies that the derivation \eqref{eq:D1b} is held, stating that HN believes UE says SUPI. Proving that the 5G-AKA-HPQC protocol provides SUPI concealment.
\end{proof}
\begin{lemma}
The 5G-AKA-HPQC protocol provides resistance against active attack by malicious SN.
\end{lemma}
\begin{proof}
The SN's initial belief \eqref{eq:D44} leads to the HN's belief \eqref{eq:D5b}, which in turn allows the SN to derive the believes \eqref{eq:D64} and \eqref{eq:D65}, related to the anchor key \(K_{SEAF}\). This sequence ensures that without \eqref{eq:D44} and \eqref{eq:D5b}, the SN cannot derive \eqref{eq:D64} and \eqref{eq:D65}, thus preventing the SN unauthorized access to anchor key \(K_{\textit{SEAF}}\). This dependency blocks a malicious SN from preemptively accessing the anchor key \(K_{\textit{SEAF}}\), proving that the 5G-AKA-HPQC protocol resists attacks by malicious SN.      
\end{proof}
\begin{lemma}
The 5G-AKA-HPQC protocol provides Perfect Forward Secrecy
\end{lemma}
\begin{proof}
Based on the derivations \eqref{eq:D3b} and \eqref{eq:D5a} and supported by the novel perfect forward secrecy axiom, it is proved that the key \( \UE\xleftrightarrow{{\langle \textit{HPK} \rangle}_{*\HN}}\HN \) achieves perfect forward secrecy.
\end{proof}

\subsection{Formal Verification of 5G-AKA-HPQC using ProVerif}

In this subsection, we carry out a rigorous protocol verification using the model-checking tool ProVerif, which is widely regarded as one of the most advanced tools for formal security analysis of cryptographic protocols. ProVerif is renowned for its ability to perform robust and comprehensive verification at the `unbounded' level, meaning it analyzes the protocol without any predefined limitations on the number of protocol executions or adversarial actions. This capability allows ProVerif to explore a vast and exhaustive range of potential attack scenarios, making it particularly suitable for identifying subtle vulnerabilities that might evade detection in more constrained analyses.

ProVerif is a state-of-the-art tool that combines advanced verification techniques with high scalability, making it capable of handling the complex requirements of modern cryptographic protocols such as the 5G-AKA-HPQC Protocol. Its efficiency in managing large protocol specifications and its ability to detect potential attack vectors through automated reasoning make it a preferred choice for security researchers. The tool generates detailed attack trace graphs that provide valuable insights into how a protocol could be exploited and the exact conditions under which these vulnerabilities occur. This information is crucial not only for diagnosing security flaws but also for guiding the refinement and enhancement of the protocol to address these weaknesses.

In this section, we document the verification process in detail, highlighting how ProVerif was employed to model the 5G-AKA-HPQC Protocol and assess its security properties. The analysis encompasses key security goals such as authentication, confidentiality, and integrity, ensuring that the protocol is evaluated against a comprehensive set of criteria. By leveraging ProVerif’s unbounded verification, we can confidently assert whether the protocol withstands adversarial attempts across all possible attack scenarios.

To enhance the reproducibility and transparency of our findings, the detailed ProVerif models used in this study have been made publicly accessible through a GitHub repository. This enables other researchers and practitioners to replicate the verification process, explore additional scenarios, or adapt the models for their own investigations. By sharing these resources, we aim to contribute to the broader community’s efforts in advancing the security of cryptographic protocols.

In conclusion, this section provides an in-depth account of the protocol verification process using ProVerif, emphasizing its strengths in scalability, comprehensive analysis, and practical insights. The results obtained through this verification process serve as a strong foundation for validating the security of the 5G-AKA-HPQC Protocol and highlight the critical role of ProVerif in modern cryptographic research.

\subsection{Queries}
The `inj-event' function was applied to ensure robust verification at each communication step, confirming the authentication and freshness of all messages transmitted in Q4, Q5, and Q6. For mutual authentication, Q4 and Q5 were modeled and analyzed to evaluate the freshness and integrity of the exchanged messages. The security of the key exchange was validated through Q1, Q2, and Q3, demonstrating that the process could be completed securely without unauthorized access by an adversary. Additionally, Q0 was employed to assess whether an attacker could access the SUPI and whether the HN correctly verified the SUCI, thereby successfully confirming the SUPI's concealment. Furthermore, Q6 was used to evaluate defenses against potential vulnerabilities stemming from compromised or malicious SNs. The phase function was integrated into the main process to verify the presence of Perfect Forward Secrecy (PFS). During this analysis, ProVerif deliberately exposed the secrets, $sk_{HN}$ and $K$, in phase 1 following phase 0 for testing purposes. Moreover, the robustness of hybrid algorithms was assessed by simulating an attacker capable of decrypting specific cryptographic schemes. This was validated by leveraging the phase function alongside Q1, Q2, and Q3 to detect any potential vulnerabilities under such conditions, as shown in table \ref{Security requirements verification query}

\begin{table}[H]

\caption{Security requirements verification query.\label{Security requirements verification query}}

\newcolumntype{M}{>{\centering\arraybackslash}m{70mm}}
\begin{tabularx}{\linewidth}{cc}
\toprule
\textbf{Security Requirment}	& \textbf{Verification Query}\\
\midrule
                                        & Q4: Inj-event(endUE\_HN\_SUPI) ==$>$ \\
                                        & inj-event(beginUE\_HN\_SUPI)\\
Mutual Authentication                                        & \&\& \\
			  	                  & Q5: Inj-event(endUE\_HN\_MAC) ==$>$\\
                                        & inj-event(beginHN\_UE\_MAC)\\

\midrule
                                         & Q4: Inj-event(endUE\_HN\_SUPI) ==$>$ \\
                                         & Inj-event(beginUE\_HN\_SUPI)\\
                                         & event(endUE\_HN\_SUPI) ==$>$ \\
                                         & event(beginUE\_HN\_SUPI)\\
                                         & Q5: Inj-event(endUE\_HN\_MAC) ==$>$\\
                                         & inj-event(beginHN\_UE\_MAC)\\
                                        & event(endUE\_HN\_MAC) ==$>$\\
                                        & event(beginHN\_UE\_MAC)\\
Availability                            & Q6: inj-event(endSN\_ANCHOR\_KEY)  \\
                                        & ==$>$ (inj-event(middleHN\_RES)\\
                                        & \quad ==$>$ ((inj-event(middleSN\_RES)\\
                                        & ==$>$ inj-event(beginUE\_RES)\\
                                        & event(endSN\_ANCHOR\_KEY)  \\
                                        & ==$>$ (event(middleHN\_RES)\\
                                        & ==$>$ ((event(middleSN\_RES)\\
                                        & ==$>$ event(beginUE\_RES)\\
                                         
\midrule
                                         & Q1: query attacker (kseafUE)\\
                                           & \&\& \\
Secure Key Exchange			  	                  & Q2: query attacker (kseafSN)\\
                                        & \&\& \\
                                        & Q3: query attacker (kseafHN)\\
\midrule
SUPI Concealment                        & Q0: query attacker (SUPI)\\
                                                                               
\midrule
Defense against                                        & Q6: inj-event(endSN\_ANCHOR\_KEY)  \\
Compromised or Malicious                         & ==$>$ (inj-event(middleHN\_RES)\\
Serving Network (SN)               & \quad ==$>$ ((inj-event(middleSN\_RES)\\
                                & ==$>$ inj-event(beginUE\_RES)\\

\midrule
Perfect Forward Secrecy                 & “phase 1; out(usch, (skHN, k)” in process\\
\midrule
                                     & “phase 1; out(usch, (skHN, k)” in process\\
                                & \&\& \\
Insecurity                              & (Q1: query attacker (kseafUE)\\
Cryptography                            & or\\
Countermeasures                             & Q2: query attacker (kseafSN)\\
                                        & or \\
                                        & Q3: query attacker (kseafHN))\\
\bottomrule 
\end{tabularx}
\end{table}

The ProVerif modeling process is divided into three primary stages: declaration, process macros, and the main process. Initially, as outlined in Algorithm \ref{declaration_queries}, we define the necessary components for protocol analysis, including types, constants, variables, channels, and security verification queries. Following this, the UE, SN, and HN are systematically represented as process macros in Algorithms \ref{UE Process Macro}, \ref{SN Process Macro}, and \ref{HN Process Macro}, respectively, allowing for a modular and reusable approach to modeling these entities.

\begin{algorithm}[H]\footnotesize
\caption{Declaration and Queries}
\label{declaration_queries}
\begin{algorithmic}[0]
\item[]	\quad \textbf{(* Channel specification *)}
\item[]	\quad free usch: channel.
\item[]	\quad free sch: channel [private]. 
\item[]	\quad 
\item[]	\quad \textbf{(* Key Encapsulation Mechanism *)}
\item[]	\quad letfun PQC\_Encaps(pubKey): bitstring, bitstring.
\item[]	\quad letfun PQC\_Decaps(bitstring, secKey): bitstring.
\item[]	\quad letfun Encaps(pubKey): bitstring, bitstring.
\item[]	\quad letfun Decaps(bitstring, secKey): bitstring.
\item[]	\quad letfun XWingKeyGen(): secKey, pubKey.
\item[]	\quad letfun XWingKeyEncap(pubKey, secKey): 
\item[]	\quad\quad\quad\quad\quad\quad\quad\quad\quad\quad\quad\quad\quad\quad\quad bitstring, bitstring.
\item[]	\quad letfun XWingKeyDecap(bitstring, bitstring, secKey): 
\item[]	\quad\quad\quad\quad\quad\quad\quad\quad\quad\quad\quad\quad\quad\quad\quad \quad\quad\quad\quad bitstring.
\item[]	\quad

\item[]	\quad \textbf{(* Secrecy verification *)}
\item[]	\quad \textbf{Q0:} query attacker (SUPI).
\item[]	\quad \textbf{Q1:} query attacker (kseafUE).
\item[]	\quad \textbf{Q2:} query attacker (kseafSN).
\item[]	\quad \textbf{Q3:} query attacker (kseafHN).
\item[]	\quad
\item[]	\quad \textbf{(* Security requirements verification *)}
\item[]	\quad \textbf{Q4:} supi: bitstring, c0: bitstring, pk: pubKey;
\item[]	\quad\quad\quad Inj-event(endUE\_HN\_SUPI(supi,c0,pk)) 
\item[]	\quad ==$>$  inj-event(beginUE\_HN\_SUPI(supi,c0,pk));

\item[]	\quad\quad\quad event(endUE\_HN\_SUPI(supi,c0,pk)) ==$>$
\item[]	\quad ==$>$  event(beginUE\_HN\_SUPI(supi,c0,pk))).
\item[]	\quad
\item[]	\quad \textbf{Q5:} supi: bitstring, k: key, rand: bitstring, sqn: bitstring;
\item[]	\quad\quad\quad inj-event(endUE\_HN\_MAC(supi, k, rand, sqn))
\item[]	\quad  ==$>$ inj-event(beginHN\_UE\_MAC(supi, k, rand, sqn));

\item[]	\quad\quad\quad event(endUE\_HN\_MAC(supi, k, rand, sqn))
\item[]	\quad  ==$>$ event(beginHN\_UE\_MAC(supi, k, rand, sqn)).
\item[]	\quad
\item[]	\quad \textbf{Q6:} supi: bitstring, k: key, rand: bitstring, sqn: bitstring;
\item[]	\quad\quad\quad inj-event(endSN\_ANCHOR\_KEY(supi, kseaf))  
\item[]	\quad ==$>$ (inj-event(middleHN\_RES(supi, k, rand, sqn))
\item[]	\quad ==$>$ ((inj-event(middleSN\_RES(res', rand)
\item[]	\quad ==$>$ inj-event(beginUE\_RES(supi, k, rand, sqn))

\item[]	\quad\quad\quad event(endSN\_ANCHOR\_KEY(supi, kseaf))  
\item[]	\quad ==$>$ (event(middleHN\_RES(supi, k, rand, sqn))
\item[]	\quad ==$>$ ((event(middleSN\_RES(res', rand)
\item[]	\quad ==$>$ event(beginUE\_RES(supi, k, rand, sqn))

\end{algorithmic}
\end{algorithm}

\begin{algorithm}[H]\footnotesize
\caption{proc\_UE}
\label{UE Process Macro}
\begin{algorithmic}

\item[] \textbf{Played by:} UE
\item[] \textbf{Input:} SUPI, sqn, $pk_{HN}$, $K$

\item[] \quad
\item[] \textbf{[Step 1.1]}
\item[] \quad let (c0, c1, c2, skue, pkue) 
\item[] \quad\quad\quad= calc\_ue\_suci(ueSUPI, pkHN) in
\item[] \quad \textbf{event beginUE\_HN\_SUPI(ueSUPI, c0, pkHN);}
\item[] \quad \textbf{out}(usch, (c0, c01, c1, c2));
\item[] \textbf{[Step 2.3]}
\item[] \quad \textbf{in}(usch, (RAND, CONC, AMF, hMAC));
\item[] \quad let ss = XWingKeyDec(RAND, skue) in
\item[] \quad let $PHK$ = KDF(ss) in
\item[] \quad let (uMAC, SQN, AK) 
\item[] \quad\quad\quad = calc\_ue\_mac(RAND, $PHK$, $K$, CONC, AMF) in
\item[] \quad if hMAC $<>$ uMAC true then exit;
\item[] \quad if hnSQN $<>$ ueSQN true then exit;
\item[] \quad \textbf{event endUE\_HN\_MAC(SUPI, $K$, RAND, ueSQN);}
\item[] \quad \textbf{event beginUE\_RES(SUPI, $K$, RAND, ueSQN);}
\item[] \quad let (RES', HXRES', $k_{AUSF}$, $k_{SEAF}$) 
\item[] \quad\quad\quad = calc\_hn\_key(RAND, $PHK$, $K$, hnSQN, AK) in
\item[] \quad \textbf{out}(usch, RES');
\item[] \quad 
\end{algorithmic}
\end{algorithm}

\vspace{-6pt}

\begin{algorithm}[H]\footnotesize
\caption{proc\_SN}
\label{SN Process Macro}
\begin{algorithmic}

\item[] \textbf{Played by:} SN
\item[] \textbf{Input:} SNname

\item[] \quad
\item[] \textbf{[Step 1.2]}
\item[] \quad \textbf{in}(usch,(c0, c1, c2));
\item[] \quad \textbf{out}(sch,(c0, c1, c2, SNname));
\item[] \textbf{[Step 2.2]}
\item[] \quad \textbf{in}(sch,(RAND, CONC, AMF, MAC, HXRES));
\item[] \quad \textbf{out}(usch,(RAND, CONC, AMF, MAC));
\item[] \textbf{[Step 2.4]}
\item[] \quad \textbf{in}(usch, RES');
\item[] \quad if HXRES $<>$ SHA((RES', RAND)) true then exit;
\item[] \quad \textbf{event middleSN\_RES(RES', RAND)};
\item[] \quad \textbf{out}(sch,(RES'));
\item[] \textbf{[Step 2.6]}
\item[] \quad \textbf{out}(sch, (SUPI, $k_{SEAF}$));
\item[] \quad \textbf{event endSN\_ANCHOR\_KEY(SUPI, $k_{SEAF}$)};
\item[] \quad
\end{algorithmic}
\end{algorithm}

\vspace{-6pt}
\begin{algorithm}[H]\footnotesize
\caption{proc\_HN}
\label{HN Process Macro}
\begin{algorithmic}

\item[] \textbf{Played by:} HN
\item[] \textbf{Input:} SUPI, $sk_{HN}$
\item[] \quad
\item[] \textbf{[Step 1.3]}
\item[] \quad \textbf{in}(sch,(c0, c1, c2, SNname));
\item[] \quad let ($k1$, $k2$) = get\_hn\_keys4supi($sk_{HN}$, c0) in
\item[] \quad if c2 $<>$ hmac($k2$, c1) true then exit;
\item[] \quad let (ueSUPI, $pk_{ue}$) = sdec(c1, $k1$) in
\item[] \quad if ueSUPI $<>$ SUPI true then exit;
\item[] \textbf{[Step 2.1]}
\item[] \quad let (ss, c1, c2) = XWingKeyEnc($pk_{ue}$, eseed)in 
\item[] \quad \textbf{event endUE\_HN\_SUPI(SUPI, c0, pk($sk_{HN}$));}
\item[] \quad get ueDB(=ueSUPI, $K$, sqn) in
\item[] \quad let $PHK$ = KDF(ss) in
\item[] \quad let SQN = calc\_sqn(sqn, c0, RAND) in 
\item[] \quad let (CONC, AK, hMAC) = calc\_hn\_mac(RAND, $PHK$, $K$, SQN, AMF) in
\item[] \quad let (XRES', HXRS', $k_{AUSF}$, $k_{SEAF}$) 
\item[] \quad = calc\_hn\_key(RAND, $PHK$, $K$, hSQN, AK) in
\item[] \quad \textbf{event beginHN\_UE\_MAC(SUPI, $K$, RAND, SQN);}       
\item[] \quad \textbf{out}(sch, (RAND, CONC, AMF, hMAC, HXRES'));
\item[] \textbf{[Step 2.5]}
\item[] \quad \textbf{in}(sch, , RES');
\item[] \quad if XRES' $<>$ RES; true then exit;
\item[] \quad \textbf{event middleHN\_RES(ueSUPI, $K$, RAND, SQN)};
\item[] \quad \textbf{out}(sch, (ueSUPI, $k_{SEAF}$));
\end{algorithmic}
\end{algorithm}

The main process, as described in Algorithms \ref{main process without fs} and \ref{main process with fs}, manages critical tasks such as cryptographic key generation and the execution of subprocesses. Algorithm \ref{main process without fs} examines scenarios where long-term keys remain secure, whereas Algorithm \ref{main process with fs} assesses the maintenance of PFS under conditions where long-term keys are compromised. Our verification emphasizes key security objectives, including mutual authentication, key exchange, availability, SUPI concealment, resilience against compromised or malicious SNs, PFS, and countermeasures against insecure cryptographic practices. To thoroughly evaluate these aspects, we define specific queries for each requirement, as presented in Table \ref{Security requirements verification query}, facilitating a comprehensive analysis of the protocol's security features.

\vspace{-6pt}

\begin{algorithm}[H]\footnotesize
\caption{Main Process without FS}
\label{main process with fs}
\begin{algorithmic}[0]
\item[] process 
\item[] \quad new $prHN$: secKey; let $puHN$ = pk($prHN$) in  
\item[] \quad new SUPI: bitstring; new $K$: key;  
\item[] \quad new SQN: bitstring;
\item[] \quad 
\item[] \quad insert ueDB(SUPI, $K$, SQN);
\item[] \quad (!UE(SUPI, $K$, SQN, $puHN$)) $|$ (!SN(SNname)) $|$ \\
\item[] \quad (!HN(SUPI, $prHN$)) $|$ phase 1; out(usch, ($prHN$, $K$))
\end{algorithmic}
\end{algorithm}

\begin{algorithm}[H] \footnotesize
\caption{Main Process without FS}
\label{main process without fs}
\begin{algorithmic}[0]
\item[] process 
\item[] \quad new $prHN$: secKey; let $puHN$ = pk($prHN$) in  
\item[] \quad new SUPI: bitstring; new $K$: key;  
\item[] \quad new SQN: bitstring;
\item[] \quad 
\item[] \quad insert ueDB(SUPI, $K$, SQN);
\item[] \quad (!UE(SUPI, $K$, SQN, $puHN$)) $|$ (!SN(SNname)) $|$ \\
\item[] \quad (!HN(SUPI, $prHN$))
\end{algorithmic}
\end{algorithm}

\subsection{Verification Results}
\begin{figure}[H]
\centering
\includegraphics[width=1.0\linewidth]{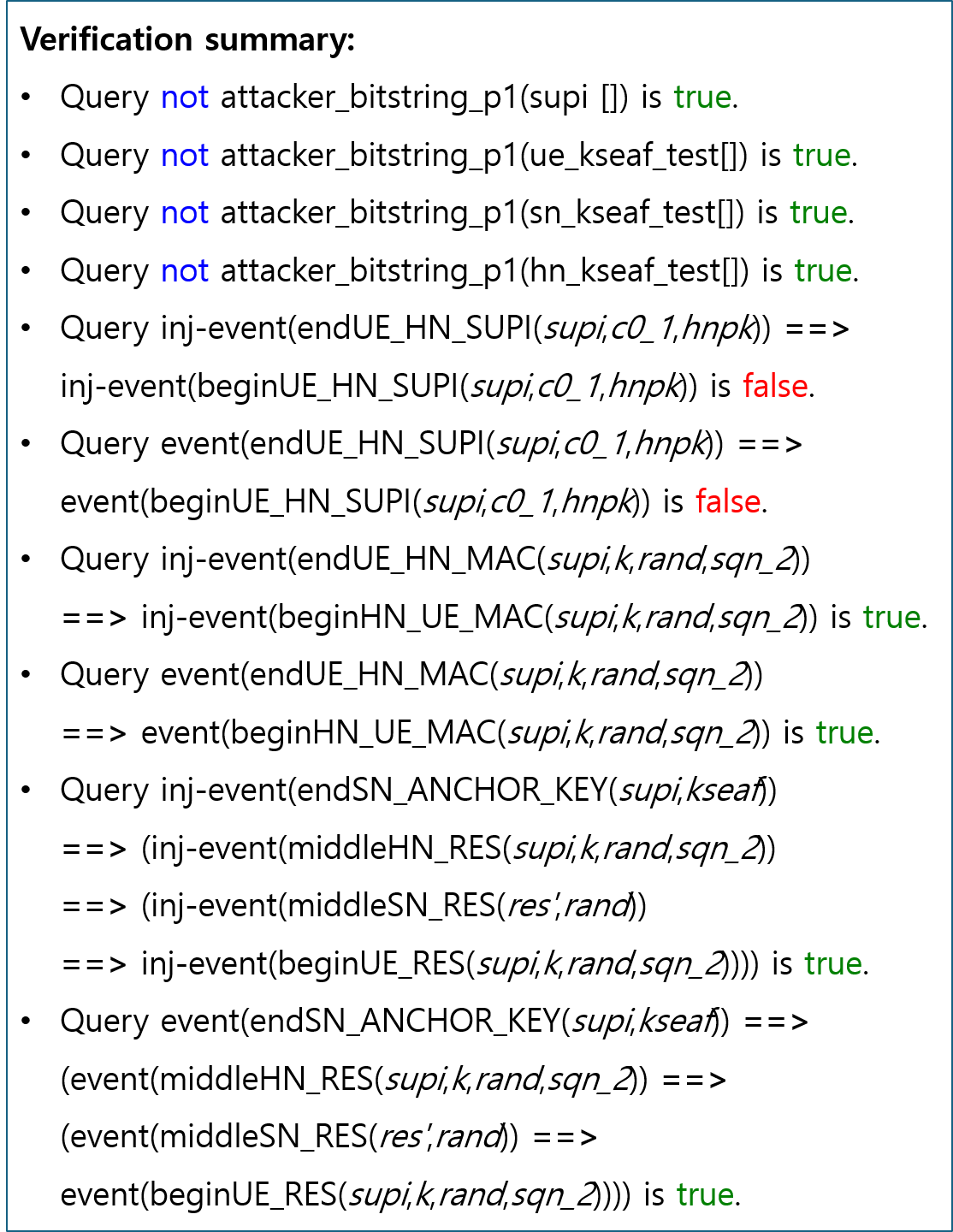}
\caption{Verification result of ProVerif of 5G-AKA-HPQC Protocol.}
\label{re0}
\end{figure}

Fig. \ref{re0} show the verification results of the 5G-AKA-HPQC Protocol, and the results are summarized in Table \ref{Attacker scenario result summary}.

\vspace{-6pt}

\begin{table}[H]

\caption{Attacker scenario result summary.\label{Attacker scenario result summary}}

\newcolumntype{M}{>{\letf\arraybackslash}m{20mm}}
\begin{tabularx}{\linewidth}{ll}
\toprule
\textbf{Attacker Scenario}	& \textbf{Result}\\
\midrule
Mutual Authentication  & True \\

\midrule
Availability  & Partially True \\

\midrule
Secure Key Exchange    & True\\

\midrule
SUPI Concealment      & True\\

\midrule
Attacks by compromised or malicious SN       & True\\

\midrule
PFS supported issue      & True\\

\midrule
Insecurity Cryptography Countermeasures       & True\\

\bottomrule 
\end{tabularx}
\end{table}

\section{Discussion}
This section primarily evaluates the security requirements of the 5G-AKA-HPQC protocol, including mutual authentication, availability, secure key exchange, SUPI concealment, defense against attacks by compromised or malicious SNs, PFS, and Insecurity Cryptography Countermeasures.

\subsection{Mutual Authentication}
The 5G-AKA-HPQC protocol is not significantly different from the 5G AKA in terms of mutual authentication between the UE  and the HN. Specifically, the UE authenticates the HN through the MAC value within the AUTN provided by the HN, and the HN authenticates the UE through the RES* value presented by the UE. More importantly, the two values used for authentication, namely MAC and RES*, are enhanced by the HPQC-based and PFS-enabled key $PHK$. This design aims to strengthen security compared to the 5G AKA protocol. According to the results of the ProVerif verification, this design allows the 5G-AKA-HPQC protocol to achieve mutual authentication between the UE and the HN in both general and PFS test cases. Moreover, Lemma 1 in the SVO logic analysis validates the protocol's mutual authentication. Consequently, we conclude that the 5G-AKA-HPQC protocol ensures mutual authentication.

\subsection{Availability}
The ProVerif verification revealed a vulnerability in the 5G-AKA-HPQC protocol regarding availability during the initial process where the UE sends the SUCI to the HN. This issue arises due to the absence of a value in the SUCI that guarantees freshness. However, this vulnerability was intentionally allowed to maintain compatibility with the existing 5G-AKA standard, which also permits this type of attack. To address this issue, improvements through mechanisms outside of cryptographic methods are required to ensure compatibility with existing technologies.

\subsection{Secure Key Exchange}
The $PHK$ is negotiated through a combination of the ECIES public key pairs of the HN and UE, along with the PQC KEM algorithm. This hybrid approach leverages the strengths of both classical and post-quantum cryptography, ensuring a robust and future-proof method for key exchange. According to the results of the ProVerif verification, the $PHK$ was securely exchanged in all test scenarios, including general test cases and PFS (Perfect Forward Secrecy) test cases, demonstrating its reliability and security under various conditions.

More specifically, the ProVerif verification highlights that the integration of the ECIES mechanism with the PQC KEM algorithm ensures the generation and exchange of keys that are not only resistant to classical cryptographic attacks but also to attacks posed by quantum adversaries. In both general test cases, where no advanced threats are simulated, and PFS test cases, which evaluate the protocol's ability to maintain forward secrecy, the $PHK$ negotiation process successfully meets security requirements. This demonstrates that the $PHK$ effectively supports the protocol’s goals of providing strong session key security, ensuring the integrity of the overall authentication and communication process between the HN and UE. 

On the other hand, Lemma 2 in the SVO logic analysis verifies the protocol’s secure key exchange by demonstrating that the UE and HN establish both direct and indirect belief in \(HPK\). Furthermore, it proves that the UE and SN gain belief in \(K_{SEAF}\), confirming that a secure key exchange has been achieved between the UE and the 5G core network (SN and HN). Therefore, we conclude that the 5G-AKA-HPQC protocol ensures secure key exchange

\subsection{SUPI Concealment}
SUPI concealment refers to the protection of a user's permanent identifier during the authentication process in 5G networks, safeguarding their privacy. SUCI is the concealed version of the SUPI, transmitted over the air to prevent the exposure of the user's identity. In the standard 5G-AKA protocol and the proposed 5G-AKA-HPQC, which supports PFS (Perfect Forward Secrecy), privacy issues arose in the process of generating SUCI if the long-term key was compromised.

However, in the case of 5G-AKA-HPQC, by adopting HPQC (Hybrid Post-Quantum Cryptography) techniques, it was demonstrated in ProVerif that SUPI could be effectively concealed, even if the long-term key is leaked, due to the robustness of the cryptographic algorithm. Moreover, Lemma 3 in the SVO logic analysis validates that the protocol provides SUPI concealment. This highlights the enhanced security of 5G-AKA-HPQC in mitigating privacy concerns associated with long-term key compromise.

\subsection{Attacks by compromised or malicious SN}
To address attacks by compromised or malicious SNs, we assume that malware could be injected into the SN, allowing it to monitor and intercept the UE's authentication messages. In the standard 5G AKA process, such malware can only access critical information like $K_{SEAF}$ and SUPI after the HN has completed authentication and shared these values with the SN. The 5G-AKA-HPQC protocol adheres to the standard and employs the same mechanism.

However, if a protocol, like DMRN ~\cite{b7}, prioritizes efficiency by delivering the anchor key and SUPI alongside the AV to the malicious SN earlier in the process, malware within the SN could immediately obtain $K_{SEAF}$ and SUPI as soon as it receives the response $RES^{*}$. This enables the attacker to access key information much earlier, exposing the system to potential breaches sooner than in the standard 5G AKA protocol.

The ProVerif analysis conducted in this study, particularly through query Q6, establishes the protocol is resilient to this threat. Moreover, Lemma 4 within the SVO logic analysis validates the protocol's robustness against active attacks initiated by malicious SNs. Consequently, it can be concluded that the 5G-AKA-HPQC protocol demonstrates a high level of security in mitigating active attacks from malicious SNs.

\subsection{PFS supported issue}
PFS is a security property that ensures the confidentiality of past communication sessions, even if the long-term keys used during those sessions are compromised in the future. The 5G-AKA-HPQC protocol employs two long-term keys: $K$ and $sk_{HN}$. To achieve PFS, it must be verified that session keys derived from these long-term keys remain secure even if the long-term keys are leaked in the future.

In this context, the critical key exchange for the negotiation of the key $PHK$, which is used to derive the master session key and anchor key, plays a pivotal role. The HPQC mechanism generates two pairs of cryptographic keys that are agreed upon using the UE's ephemeral PQC public key $pk_{UE}$. Importantly, the corresponding private key $sk_{UE}$ is deleted immediately after the protocol execution, making it irrecoverable in the future.

As a result, PFS is preserved for the key $PHK$ and the session keys derived from it. This ensures that even if the long-term keys $K$ or $sk_{HN}$ are compromised in the future, the confidentiality of past sessions remains intact, maintaining the protocol's resilience against retrospective attacks. Furthermore, Lemma 5 in the SVO logic analysis provides additional validation of these findings, confirming that the \(PHK\), serving as the key material for the anchor key \(K_{SEAF}\), effectively ensures the achievement of perfect forward secrecy.

\subsection{Insecurity Cryptography Countermeasures}
Since PQC is still in the process of standardization, testing in real-world environments has not yet been conducted. As a result, PQC algorithms may pose potential vulnerabilities, making it essential to combine them with traditional cryptographic methods. The IETF is standardizing technologies such as HPKE~\cite{b8} and X-Wing to address these concerns. Among them, the 5G-AKA-HPQC protocol adopts X-Wing, which provides stronger security, to mitigate potential vulnerabilities.

In ProVerif, the adversary was configured to decrypt using either the PQC-based cryptography or the traditional ECIES-based cryptography, testing the extent to which an attacker could compromise the protocol. The verification results demonstrated that the 5G-AKA-HPQC protocol remains secure under these conditions.

\section{Conclusions}
In this paper, we proposed an enhanced 5G-AKA protocol, referred to as 5G-AKA-HPQC, which addresses vulnerabilities posed by potential quantum threats while ensuring UE unlinkability, quantum-safe security, and  
perfect forward secrecy. The proposed protocol is compatible with the 3GPP standard and incorporates the robust HPQC algorithm, X-Wing, which is emerging as a strong candidate in the ongoing standardization efforts by IETF alongside HPKE. Given the superior cryptographic strength of X-Wing, it has been adopted as the core cryptographic mechanism in the proposed protocol.

The validity of the 5G-AKA-HPQC protocol was rigorously verified using formal methods such as SVO Logic and ProVerif. Furthermore, a comparative analysis was conducted to evaluate the security properties and computational/communication overheads of 5G-AKA-HPQC against existing protocols, including those standardized by 3GPP (5G-AKA)~\cite{b9} and IETF (EAP-AKA`)~\cite{b10}. The analysis demonstrated that the proposed protocol effectively balances enhanced security with computational and communication efficiency.

As 5G-AKA and EAP-AKA` remain the cornerstone security protocols in 3GPP and IETF standards, respectively, this study provides a foundation for extending our research to improve security protocols within the EAP framework. Future work will involve proposing and implementing a quantum-safe enhanced security protocol for the EAP framework by integrating HPQC mechanisms, followed by a comprehensive performance evaluation to benchmark its efficiency and security.

\begin{center}
Biography    
\end{center}

\begin{IEEEbiography}[{\includegraphics[width=1in,height=1.25in,clip,keepaspectratio]{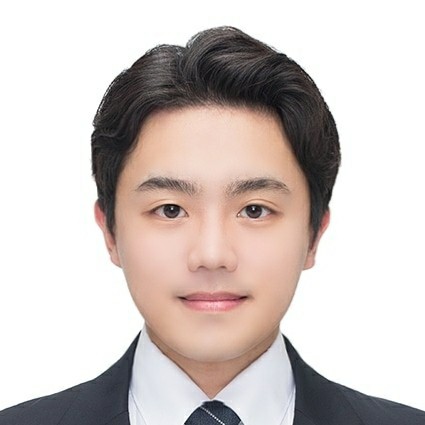}}]{Yong ho Ko} received the B.S. and M.S. degrees in infromation security from Soonchunhyang University, Asan, South Korea, in 2018, 2020, separately. He is currently working as an Ph.D. graduate student with the Department of Financial Information Security, Kookmin University, Seoul, South Korea. His research interests include the 5G/6G, drone security, and formal security verification.
\end{IEEEbiography}

\begin{IEEEbiography}[{\includegraphics[width=1in,height=1.25in,clip,keepaspectratio]{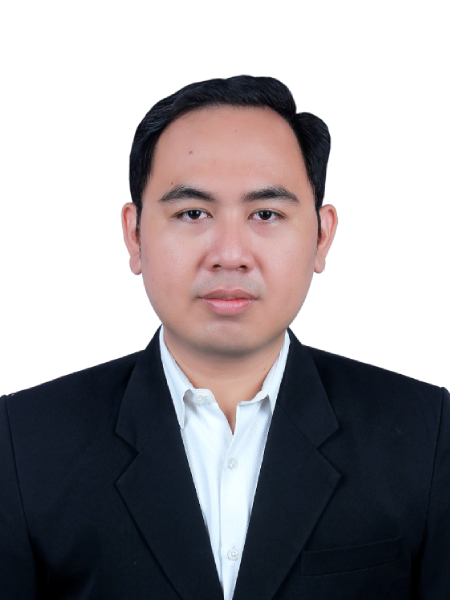}}]{I Wayan Adi Juliawan Pawana} (Member, IEEE) received the S.Kom. degrees in computer science and M.T. degree in electrical engineering from Udayana University, Bali, Indonesia in 2014
and 2021, respectively. He is currently working as a Ph.D. graduate student at the Department of Financial Information Security, Kookmin University, Seoul, South Korea. His research interests include the 5G/6G networks, internet security, and AI security.
\end{IEEEbiography}

\begin{IEEEbiography}[{\includegraphics[width=1in,height=1.25in,clip,keepaspectratio]{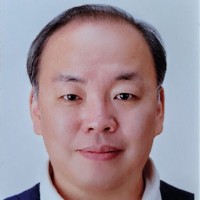}}]{Ilsun You} (Senior Member, IEEE) received the M.S. and Ph.D. degrees in computer science from Dankook University, Seoul, South Korea, in 1997 and 2002, respectively, and the Ph.D. degree from Kyushu University, Japan, in 2012. He is currently working as a Full Professor with the Department of Financial Information Security, Kookmin University, South Korea. His research interests include internet security, authentication, access control, and formal security analysis. He is a fellow of the IET. He is on the Editorial Board of Information Sciences, Journal of Network and Computer Applications, International Journal of Ad Hoc and Ubiquitous Computing, Computing and Informatics, and Intelligent Automation and Soft Computing.
\end{IEEEbiography}

\vfill

\end{document}